# Local-ring network automata and the impact of hyperbolic geometry in complex network link-prediction

Alessandro Muscoloni[1], Umberto Michieli[1,2] and Carlo Vittorio Cannistraci[1,3,*]

[1]Biomedical Cybernetics Group, Biotechnology Center (BIOTEC), Center for Molecular and Cellular Bioengineering (CMCB), Center for Systems Biology Dresden (CSBD), Department of Physics, Technische Universität Dresden, Tatzberg 47/49, 01307 Dresden, Germany
[2]Department of Information Engineering, University of Padova – Via Gradenigo, 6/b, 35131 Padova, Italy
[3]Brain bio-inspired computation (BBC) lab, IRCCS Centro Neurolesi "Bonino Pulejo", Messina, Italy

*Corresponding author: kalokagathos.agon@gmail.com

## Abstract

Methods for topological link-prediction are generally referred as *global* or *local*. The former exploits the entire network topology, the latter adopts only the immediate neighbourhood of the link to predict. Global methods are 'believed' to be the best performing. Is this common belief well-founded?
Stochastic-Block-Model (SBM) is a global method believed as one of the best link-predictors and widely accepted as reference when new methods are proposed. But, our results suggest that SBM, whose computational time is high, cannot in general overcome the Cannistraci-Hebb (CH) network automaton model that is a simple local-learning-rule of topological self-organization proved by multiple sources as the current best *local-based* and parameter-free *deterministic* rule for link-prediction. In order to elucidate the reasons of this unexpected result, we formally introduce the notion of local-ring network automata models and their tight relation with the nature of common-neighbours' definition in complex network theory.
In addition, after extensive tests, we recommend Structural-Perturbation-Method (SPM) as the new best global method baseline. However, even SPM overall does not outperform CH and in several evaluation frameworks we astonishingly found the opposite. In particular, CH was the best predictor for synthetic networks generated by the Popularity-Similarity-Optimization (PSO) model, and its performance in PSO networks with community structure was even better than using the original internode-hyperbolic-distance as link-predictor. Interestingly, when tested on non-hyperbolic synthetic networks the performance of CH significantly dropped down indicating that this rule of network self-organization could be strongly associated to the rise of hyperbolic geometry in complex networks.
In conclusion, we warn the scientific community: *the superiority of global methods in link-prediction seems a 'misleading belief' caused by a latent geometry bias of the few small networks used as benchmark in previous studies*. Therefore, we urge the need to found a latent geometry theory of link-prediction in complex networks.

# 1. Introduction

The aim of topological link-prediction is to detect, in a given network, the non-observed links that could represent missing information or that may appear in the future, only exploiting features intrinsic to the network topology. It has a wide range of real applications, like suggesting friendships in social networks or predicting interactions in biological networks [1]–[3]. Although this study is focused on monopartite networks, link-prediction has recently been successfully implemented also in different types of network topologies such as bipartite [4], [5] and multilayer networks [6].

The link-prediction methods, according to the type of topological information exploited, can be broadly classified in two main categories: global and local. Global methods take advantage of the entire network topology in order to assign a likelihood score to a certain non-observed link. On the contrary, local approaches take into consideration only information about the neighbourhood of the link under analysis [1], [3].

In 2009, Guimerà et al. proposed a new global inference framework based on stochastic block model (SBM) in order to identify both missing and spurious interactions in noisy network observations [7]. The general idea of a block model is that the nodes are partitioned into groups and the probability that two nodes are connected depends only on the groups to which they belong. The framework introduced is a global approach where, assuming that there is no prior knowledge about which partition is more suitable for the observed network, the likelihood of a link can be computed theoretically considering all the possible partitions of the network into groups. Since this is not possible in practice, the Metropolis algorithm, which is based on a stochastic procedure, is exploited in order to sample only a subset of partitions that are relevant for the estimation of the link reliability [7]. The high computational time becomes anyway prohibitive for large networks and restricts the range of applicability to small networks up to at most a few thousand nodes [1].

However, in many recent link-prediction studies where new methods are proposed, SBM is considered among the best state-of-the-art methods to adopt as a baseline for a performance comparison. For the sake of clarity, here we explicitly report a few examples.

From [8]: *"Roger Guimerà et al. proposed a Stochastic Block Model (SBM) which can predict both missing links and spurious links and is able to give much better accuracies of prediction on various kinds of networks than current popular methods including the HRG approach [33]. Because the SBM algorithm is a state-of-the-art approach which has very outstanding accuracy performance of link prediction on undirected networks without additional node's or edge's* *attribute information, we mainly make performance comparisons on both accuracy of missing link prediction and computational efficiency between our algorithm and the SBM approach."*

From [9]: *"Surprisingly, by directly applying the first-order matrix perturbation method, we achieve more-accurate link predictions than some gracefully designed methods such as HSM [18] and SBM [25]."*

From a recent survey on the link-prediction state-of-the-art [10]: *"Stochastic block model, characteristics: outperforms at identifying both missing links and spurious links; high computation time. [...] probabilistic graph models achieve better performance than basic topology-based metrics, especially improve the prediction accuracy."*

Other investigations that consider SBM a state-of-the-art global method are [11]–[15].

Link-prediction studies are not always convincing and exhaustive in the selection of the approaches to adopt as a reference for comparison. Furthermore, the malpractice to consider few (a tenth) small-size (less than 1000 nodes) networks as benchmark can bring to wrong conclusions. For honesty, we report that also one publication [3], led by the last author of this study, was characterized by this issue.

Two observations triggered our attention. Firstly, the remarkably high computational time of SBM and the consequent network size constraint for its application. Secondly, the fact that, as far as we are concerned, we could not find in the scientific literature convincing proofs of SBM's (and other global methods) superiority with respect to the best local methods. Moving from these premises, we decided to conduct an accurate study that compares SBM and the most promising algorithms for link-prediction. Hence, in this article, we made a thoughtful analysis of the best global and local methods, which we extensively (about 30 real networks used plus different models of artificial networks) tested on 4 different evaluation frameworks both with small (less than 1000 nodes) and large (from 3000 up to 40000) networks. We considered evaluations on re-prediction of random removed links and on network evolution across time. We compared performance on real and artificial networks. We evaluated also the overlap and diversity between the true links predicted by global and local methods. After the study of Liben-Nowell and Kleinberg [2], this represents the largest and most recent study on testing state-of-the-art methods for topological link-prediction in complex unweighted and undirected monopartite networks. We believe that, at this stage, the link-prediction field needs a reorganization of the knowledge in order to reach an agreement and set clear guidelines for the future. Here, we attempt to propose the baseline methods and the evaluation strategies that, for a fair comparison, should be taken into consideration and included in forthcoming studies. In addition, we introduce the new concept of local-ring network automata and investigate their impact to model network geometry and prediction of connectivity (links) in real complex networks.

## 2. Results

**2.1 Introducing the CH model and the general idea of local-ring network automata for link-prediction**

The Cannistraci-Hebb (CH) network automaton model is a simple local-learning-rule of topological self-organization. Given a network as framework, CH is a network automaton able to grow (learn) new links between the nodes following a certain paradigm. The theory behind this mechanistic (physical) model of network self-organization is the *local-community-paradigm* (LCP). Initially detected in brain-network self-organization topology [3] and afterward extended to any monopartite and bipartite [4] complex network, the LCP theory derives from a purely topological inspired interpretation of a local-learning-rule of neuronal networks named Hebbian learning rule [3]. A thoughtful analysis of the fundaments that advocate this theory is offered in a dedicated paragraph of the Discussion section. At this point, we just need to report that one of the corollary of the LCP theory suggests that neighbourhood-based (local-based) topological link-prediction should complement the information content related with common neighbour nodes using also the topological knowledge emerging from the cross-interactions between them. In the past publications the CH model was referred to as Cannistraci-Resource-Allocation (CRA) rule [3]–[5], [8], [14], [16]–[20] and was interpreted only as a local, parameter-free and model-based deterministic rule for topological link-prediction in both monopartite [3] and bipartite networks [4], [5]. However, we feel that this interpretation was too technical and, to a certain extent, also misleading.

In this article we want to present that CH and the other local, parameter-free and model-based deterministic rules for topological link-prediction are much more than simple methods for link-prediction, and we consider this message a main conceptual innovation of this study. They are in reality network automata that participate to learn and form new structures on a network by growing according to some local-rules, principia, mechanisms of self-organization.

At this point, before proceeding with more detailed explanations, we need to provide some definitions of basic concepts that will be extensively used in the next paragraphs.

Given two nodes *i* and *j*:

- *Local-path*: the smallest possible path allowed between *i* and *j* on a certain network topology, assuming *i* and *j* non-adjacent. For instance: in monopartite networks the local-path is a two-steps path, whereas in the bipartite networks it is a three-steps path.
- *Local-loop*: a local-path closed by a link between *i* and *j*.
- *Common-neighbours (CNs)*: the nodes involved in all the local-loops, *i* and *j* excluded.
- *Local-community-links (LCLs)*: links between the CNs.
- *Local-community*: subnetwork composed by the CNs and the LCLs.
- *External-local-community-links (eLCLs)*: CN links that are external to the local-community and are not directed to the nodes *i* and *j*.
- *Local-tunnel*: the topological structure created by the ensemble of all the local-paths between *i* and *j*, and the LCLs between CNs.
- *Local-ring:* a local-tunnel closed by a link between *i* and *j*.
- The volume of the local-tunnel or local-ring is proportional to the number of CNs.
- The existence of a local-tunnel or local-ring is maximized by the number of LCLs and minimized by the number of eLCLs.

Based on these definitions, we discuss as a first example the common-neighbours (CN) link-prediction index (also named common-neighbours similarity) and our reinterpretation as a network automaton. The generative mechanism of self-organization of the CN network automaton model is to prioritize the appearance of the missing link which *maximizes the number of nodes involved in the local-ring* that it would close. In practice, it prioritizes the formation of the missing link that transforms the largest local-tunnel in the network in a local-ring. In a monopartite topology, the local-path between two nodes consists of two steps and a local-loop is a triangle. Therefore, the previous CN general rule implies that the link with the highest likelihood to appear is the one that *maximizes the number of nodes involved in the ensemble of triangles* that it would close (triadic closure mechanism). We let notice that in the monopartite case the number of nodes involved in the triangles (local-loops) is equivalent to the number of triangles itself. This led to the common idea that the CN mechanism is defined according to the triadic closure [21], [22], since it prioritizes the links that maximize the number of triangles between two non-adjacent nodes. However, we warn that this is true only in the monopartite case, and it would not be correct to generalize the CN mechanism for other network topologies stating that it prioritizes the new-forming links that *maximize the number of local-loops* between two non-adjacent nodes. This misleading idea prevented for many years the definition of common-neighbours on bipartite networks [22]–[25], since the triadic closure cannot be defined, and the problem was recently solved by Daminelli et al. [4]. In a bipartite topology, the local-path between two nodes of different classes consists of three steps and a local-loop is a quadrangle. Therefore, the CN general rule implies that the link with the highest likelihood to appear is the one that *maximizes the number of nodes involved in the quadrangles* that it would close (quadratic closure mechanism) [4], [5]. In addition, we clarify that in the bipartite case the number of quadrangles is not equivalent to the number of nodes involved in the quadrangles, but to the number of LCLs [4], [5]. In the case of the CH network automaton model the likelihood to appear of a new link is function not only of the CNs, but also function of the LCLs. Since this is in practice a generalization and reinterpretation of a Hebbian learning local-rule to create new topology in networks, we decided to rename it as Cannistraci-Hebbian (CH) network automaton model, rather than Cannistraci-Resource-Allocation (CRA) as named in the previous articles [3]–[5], [8], [14], [16]–[20].

Fig. 1 suggests a geometrical interpretation about how the CH network automaton model works in a monopartite topology. The *local-tunnel* (which is formed in a hidden high-dimensional geometrical space that here, for simplifying the visualization, we project in the hyperbolic disk) provides a route of connectivity between the two non-adjacent nodes. Although in the hyperbolic disk visualization the local-tunnel has a shape that resembles the hyperbolic distances (curved toward the centre), we clarify that the proposed term, tunnel, is only an idealized definition; in fact the shape of this bridging structure could also geometrically look as a chamber or basin, but this does not change the meaning of the definition.

The higher the number of CNs, the higher the volume of the local-tunnel. For each CN, the higher the number of LCLs in comparison to the eLCLs, the more the shape of the local-tunnel is well-defined and therefore its existence confirmed. Therefore, in link-prediction, CH estimates a likelihood that is proportional both to the volume of the local-tunnel and to the extent to which the local-tunnel exists. On the other hand, the CN network automaton model only estimates a likelihood proportional to the volume of the local-tunnel, which is a significant limitation. The formula of the CH network automaton model in a monopartite network, together with an example of how to compute the likelihood to grow a new link, are given in Fig. 2. In brief - defining the local-community as the cohort of cross-interacting common-neighbours - CH represents a weighted version of the common-neighbours index, where each common neighbour is penalized for links external to the local-community and rewarded for links internal to the local-community. The local-community represents the central chunk of the local-tunnel exclusively formed by CNs and LCLs. Accurate tests on several real networks in both monopartite [3], [8], [14], [18], [19] and bipartite [4], [5], [17] topologies confirmed the theory's validity, in fact the LCP-based variations of the standard CN-based link predictors showed a significant improvement. Several articles have pointed out the effective prediction capabilities of CH [3], [8], [14], [18], [19] and two recent extensive studies [16], [20] have confirmed that up today it seems actually the best performing state-of-the-art local approach for topological link-prediction.

Finally, we clarify that this article is the first study that introduces the idea, terminology and definitions to explain the network-automaton-driven local-growth of new links in complex networks. It presents a general formalization in terms of *local-ring closing procedure* and, most importantly, in terms of what this closing procedure is involving in the local-ring (more common neighbour nodes or more links between the common neighbour nodes). We can conclude that the formalization of network automata here provided is fundamentally dependent from the definition and formalization of the local-ring concept in complex networks. Hence, we call the class of network automata here introduced: *local-ring network automata for link-prediction*. The definition of local-ring is dependent from the definition of local-loop that, in turn, is dependent from the type of network topology: monopartite, bipartite, etc. Other types of network automata mechanisms were already defined on complex networks [26]–[28] but, to the best of our knowledge, we exclude that they were directly linked to the effective task to perform link-prediction in general, and to the local, parameter-free and model-based deterministic rules for topological link-prediction in particular. In brief, the field of network automata was till now explored by mathematicians and had a low impact on real applications. In this study, we demonstrate that local-ring network automata might have instead a dramatic impact to model network geometry and prediction of connectivity in real complex networks. Hence, we endorse the intuition (till now not supported by positive evidence in applied fields) of Stephen Wolfram [29] on the importance of network automata to model and predict the structural evolution of complex systems. His intuition was maybe ante-litteram, and we hope that the results provided in the next link-prediction evaluations might help to gain interest in the practical applications of local-ring network automata on real complex big data.

## 2.2 State-of-the-art methods for link-prediction

In this link-prediction investigation we decided to focus the attention on 3 state-of-the-art approaches to be compared with SBM. They are based on completely different theories that will be now concisely introduced, together with the explanation of their choice (for further details please refer to the Methods section).

The first method we considered is the Structural Perturbation method (SPM), a global approach that relies on a theory similar to the first-order perturbation in quantum mechanics [9]. It implements a perturbation procedure based on the idea that a missing part of the network is predictable if it does not significantly change the structural features of the observable part, represented by the eigenvectors of the matrix. Therefore, assumed the perturbed matrices to be good approximations of the original adjacency matrix, they are exploited for assigning likelihood scores to the non-observed interactions [9]. The original publication [9] already suggested SPM to be a promising method able to clearly outperform SBM. A recent study [20] has confirmed that SPM is actually one of the best performing state-of-the-art global approaches for topological link-prediction.

The second method we considered is the CH network automaton model and was introduced in the previous section.

The third method we considered is Fast probability Block Model (FBM), a global method based on the same network partitioning theory as SBM, but it replaces the Metropolis algorithm introducing a greedy stochastic strategy for an efficient sampling over the space of the possible partitions, which leads to high improvements in the computational time [8].

To sum up, we focused our choice on the two methods that recent studies have demonstrated to be the best performing respectively for the global and the local approaches, and as third one we considered a faster variant of SBM. The four methods have been tested on both artificial and real complex networks and the results will be now discussed.

## 2.3 Evaluation on small-size real complex networks

In order to compare the performance of the link-prediction methods analysed in this study, a dataset of 19 small-size networks has been collected from different real-world domains. Due to the computational time constraints imposed by SBM, only networks of size up to around one thousand nodes have been considered. Several statistics of the real networks are shown in Table 1. The dataset is intended to cover topologies having as much as possible different characteristics, in order to avoid to favour methods tending to perform better in presence of particular structural properties.

When no information is available about missing or future interactions, the standard procedure adopted for evaluating the link-prediction performance on a given network is the following: (a) a certain number $r$ of links are randomly removed from the network; (b) the algorithm is executed in order to obtain a ranking of the non-observed links in the reduced network by decreasing likelihood scores; (c) the precision is computed as the percentage of removed links among the top-*r* in the ranking; (d) the previous steps are repeated for several iterations and the average precision is reported as measure of performance for the algorithm on the given network. A common and accepted practice that we have also adopted is to set *r* equal to 10% of the links in the network. The value of 10% removal is commonly accepted [3], [9] because it is proven to generate missing interactions in the network without significantly affecting the main topological properties. Larger removal percentages can cancel important topological information such as local-community organization [3]. For the methods SPM, CH and FBM the evaluation procedure has been repeated for 100 iterations, whereas for SBM it has been limited to 10 iterations due to the high computational time.

Another evaluation metric widely used is the area under the ROC curve (AUC-ROC or, short name, just AUC). However, a recent comprehensive investigation focused on the problem of link-prediction evaluation [30], followed and supported by successive link-prediction studies [4], [20], has pointed out how AUC can be deceptive and it has strongly advised against the adoption of this metric. The first reason is that AUC should be used for the evaluation of a classification problem in which a positive and a negative set are present, but in a link-prediction problem it is not appropriate to consider the non-observed links as negative links, therefore a negative set cannot be well defined. Secondly, even if it were considered a classification problem, it would be characterized by an extreme imbalance between the positive and negative sets, since most of the real networks are sparse. In this situation ROC curves and their areas fail to honestly convey, represent and quantify the difficulty of the prediction problem, leading to exceptionally high scores even when the precision would be particularly low [30], we therefore further discourage its usage.

Table 2 reports the precision evaluation of the four methods for each real network. The maximum level of precision reached on the different networks is quite variable, going from 0.08 in *physicians innovation* up to 0.84 in *littlerock foodweb*. Looking at the best methods for each network, highlighted in bold, it is evident that SPM obtains the highest performance in 11 out of 19 networks, sharing the first position in two of them, whereas SBM reaches the best prediction in only two networks, with one first position shared. The gap between the best and worst method for each network is in general contained within a level up to 0.2, however, a few outliers can be noticed. The first one is *netsci* with a divergence of 0.37 between CH, the best method, and FBM, the worst. But much more relevant is the case of *littlerock foodweb*, where it occurs an atypical discrepancy between SPM-SBM, respectively 0.84 and 0.73 of precision, versus FBM-CH, with 0.17 and 0.15 of precision. This network offers a clear example for disproving the suitability of the mean precision as an overall metric of best performance across several networks. The mean precision, even if reported for the sake of completeness, is particularly sensitive to the presence of such networks in which certain topological properties favour the prediction only for some methods, creating a huge gap between the various performances. The introduction of a few of this kind of networks would certainly bias the comparison toward the methods that fit with them. In fact, if we compute the mean precision excluding *littlerock foodweb*, as shown in Table 2, the order of overall performance for SBM and CH would be inverted with respect to the one that includes the network. However, it has to be clear that we are not suggesting to exclude in future studies peculiar networks that lead to an anomalous divergence in performance between the methods, in fact, they still represent real-world topologies and for a fair comparison the dataset should be as rich and diverse as possible. On the contrary, we want to encourage to include such networks and to use a more robust and reliable metric for assessing the overall performance, in order to establish a final ranking of the methods.

The evaluation we propose, already adopted in two recent link-prediction studies [20], [31], is the precision-ranking. After the computation of the precision as previously described and shown in Table 2, the methods are ranked for each network by decreasing precision, considering an average rank in case of ties. The mean ranking of the methods over all the networks represents the final evaluation score, the values are reported in Table 3. The best performing approach, as already deducible from Table 2, results to be SPM, with an average ranking of 1.89. The second method is CH with 2.50, meaning that it ranked on average around half a position lower than SPM. The third approach is SBM with 2.71 and, as last, FBM, which ranked on average exactly one position lower than SPM.

Similarly to Table 2, Table 3 reports also the mean score computed without considering *littlerock foodweb*. The introduction of the ranked values has attenuated the big gap of performance in the network and consequently the final scores appeared to be robust. This is actually the goal of the precision-ranking evaluation, when multiple methods are compared across several networks it prevents that a unique but consistent alteration in the set of networks will substantially subvert the overall evaluation.

In order to check the statistical significance of the difference in performance between the methods, pairwise permutation tests (10000 iterations) for the mean have been performed using for each method the precision-ranked values over the networks (columns of Table 3). Table 4 reports for each pair of methods the p-value of the test, adjusted for multiple hypothesis comparison by the Benjamini–Hochberg correction. Considering a significance level of 0.05, the only pairs whose mean performances are significantly different are SPM-SBM and SPM-FBM, which are actually the ones with a difference in the mean ranking corresponding to around one ranking-position. This result corroborates the classification of SPM as one of the best state-of-the-art global methods. It has to be noticed that CH, the only local approach, obtained an overall score of performance higher (although not significantly) than two global methods like SBM and FBM, and its gap with SPM is not statistically significant. This is the first important finding of this article and confirms the result showed in previous studies about the effective prediction capabilities of CH, despite exploiting a restricted amount of topological information with respect to the other approaches here considered.

Table 5 reports the computational time required by the methods in order to perform the link-prediction. The algorithms have been run in the same workstation, for further details please refer to the Hardware and software section. The table highlights that SPM, CH and FBM are quite fast as methods and respectively required from a few seconds up to slightly more than one minute for the link-prediction on a network of around one thousand nodes. SBM, instead, required around 2 days for the same task. This table, together with Table 4, advocates the second crucial finding of this study: SBM displays huge computational time in comparison to the best state-of-the-art approaches, without an overall significant gain in link-prediction performance even versus the best local method. For a deeper investigation, the correct predictions shared by the different methods have been also analysed. For each small-size real network (and for each of 10 iterations), considering the entire set of links that have been correctly predicted by a pair of methods, the percentage of these links that are shared or not between the two methods is computed. The mean of the percentages taken over all the networks and iterations are reported as a Venn diagram for each pair of methods in Suppl. Fig. 1.

It is possible to notice that on average the different approaches share around half of the correctly predicted links. The remaining part is distributed in an almost balanced way, with a few percentage points more for the better performing method among the two. The couple having the highest overlap is CH-FBM with 62% and, although based on the same theory, the one with the lowest overlap is SBM-FBM.

From this last analysis emerges that a proper combination of even only two of these methods would significantly increase the number of correctly predicted links. Therefore, in order to exploit this link-prediction heterogeneity across methods, we advance the idea to build hybrid methods that should potentially lead to higher performances.

## 2.4 Evaluation on artificial networks

We were surprised to discover that a local method such as CH is comparable (because its performance is not significantly different) to SPM - that is the best global method. Possibly, the previous results could have been only part of the picture and biased by the selection of small-size real networks available in literature, therefore we decided to extend the link-prediction evaluation considering also artificial networks.

The Popularity-Similarity-Optimization (PSO) model is a network model recently proposed [32], which explains the complex networks growth according to a procedure that optimizes the trade-off between nodes popularity and similarity. The hyperbolic space offers an adequate representation for this evolution process, where the radial coordinate and the angular distance are the geometrical counterparts of the nodes popularity and similarity [32]. Here, the PSO model has been used to generate artificial networks with parameters $\gamma = 3$ (power-law degree distribution exponent), $m = [10, 12, 14]$ (half of average degree), $T = [0.1, 0.3, 0.5]$ (temperature, inversely related to the clustering coefficient) and $N = [100, 500, 1000]$ (network size). The values chosen for the parameter $m$ are centered around the average $m$ computed on the dataset of small-size real networks. The values chosen for $N$ and $T$ are intended to cover the range of network size and clustering coefficient observed in the dataset of small-size real networks in Table 1. Since the average $\gamma$ estimated on the dataset of small-size real networks is higher than the typical range $2 < \gamma < 3$ [33], we choose $\gamma = 3$.

Fig. 3 reports for each parameter combination the average link-prediction precision computed on 100 networks for SPM, CH and FBM, and on 10 networks for SBM, due to the high computational time. The first fact to highlight is that the methods obtain different performances for low temperature (high clustering), similar performances for medium temperature (medium clustering) and almost the same performance for high temperature (low clustering). Furthermore, they all exhibit a decreasing behaviour going from low to high temperature. This is expected since, according to the PSO model theory, for increasing temperature the network tends to assume a more random and degenerate topology, which makes the link predictability harder.

Focusing on the low temperature $T = 0.1$, CH, which is the only local approach, outperforms the other global methods for all the combinations of $N$ and $m$, followed by SPM. SBM surpasses FBM only for networks of size $N = 100$, whereas it is the worst performing for networks of bigger size. For higher temperatures, even if the performance difference becomes thinner and sometimes vanishes, the same trend is preserved or at least not significantly inverted.

Table 6 reports the mean ranking of the methods over all the parameter combinations of the PSO model and, analogously to Table 4, the pairwise p-values of the permutation test (10000 iterations) for the mean ranking. As already suggested by the precision plots, CH obtains the best mean ranking with 1.30, significantly higher than all the other approaches. SPM is in second position with 1.78, still significantly higher than the two other global methods. FBM is slightly better than SBM, although not significantly, and they obtain a score more than two ranking-positions far from CH.

The ranking of the methods is not exactly the same as in the small-size real networks dataset, which - we speculate - it might suggest that, although some structural properties are reproduced by the model, it does not cover the whole variability present in the real network topologies. Conversely, it might be true also the opposite, that the selection of real complex networks we used is biased towards network topologies that favour global models, while the artificial networks not. However, the two separate evaluations are still in agreement on one point: the two methods that recent studies have demonstrated to be among the best performing for the global and the local approaches, respectively SPM and CH, obtained a higher overall performance with respect to the two methods based on the stochastic block model theory, SBM and FBM.

Since SBM might be sensitive to the organization of the network in blocks and these artificial networks do not have communities, we repeated the same simulations on artificial networks generated using the nonuniform PSO (nPSO) model, a variation of the original PSO model introduced in order to confer an adequate community structure [34]. Suppl. Fig. 2 shows examples of networks comparing the original PSO model and the nPSO model with 4 and 8 communities. From the link-prediction performance in Fig. 4 it is possible to see that, while for networks of size $N = 100$ the ranking of the methods is variable and the precisions are on average comparable, for networks of increasing size CH tends to significantly ($p<0.05$) surpass the global methods. Overall Table 7 underlines that, even with the introduction of the communities, CH obtains the best mean ranking with respect to the global approaches in a significant way. Furthermore, we can notice that decreasing the number of communities does not alter the picture, as demonstrated in Suppl. Fig. 3 where 4 communities are adopted. This suggests that the hyperbolicity of the networks might be the main cause of this result and this point will be better analysed in the Discussion section. As additional comment, we let notice that the nPSO model captured the quite comparable performance of the methods on the smallest networks as observed on real topologies, offering a more realistic framework with respect to the original PSO model. Finally, using the nPSO is evident the gain of performance of CH for networks of size 500 and 1000 nodes. This result suggests that, if the nPSO is well-designed to be realistic, CH should outperform also the other methods in link-prediction on large size real networks with hyperbolic geometry.

The plots in Fig. 3, Fig. 4 and Suppl. Fig. 3 report also as a reference the performance that is obtained if the links are predicted ranking them by the hyperbolic distances (HD) between the nodes in the original network. It can be noticed that in the PSO model the performance is slightly lower than CH for $T = 0.1$ and higher than CH for $T = [0.3, 0.5]$, whereas in the nPSO model, with the introduction of the communities, the HD performance consistently decreases. Since the real networks tend to present a community structure, this result suggests that embedding a network in the hyperbolic space and using the ranked HD for predicting the links will not generally lead to high values of precision. Furthermore, we let notice that in the generative procedure of the PSO model the links are not always established between the closest nodes, but with a probability dependent on the hyperbolic distance, therefore the usage of the HD-ranking might be not the best solution for link-prediction.

## 2.5 Evaluation on time-evolving real networks

In the wake of the results obtained on the artificial networks, where the local CH model outperformed the global models especially for networks of increasing size, we were encouraged to look for further clues using also large-size networks and a different evaluation framework. Therefore, we here propose an investigation that considers the link-growth evolution of a real network over time.

The networks represent six Autonomous systems (AS) Internet topologies extracted from the data collected by the Archipelago active measurement infrastructure (ARK) developed by CAIDA [35], from September 2009 to December 2010 at time steps of 3 months. Several statistics of the AS snapshots are shown in Table 8, they are large-size networks with a number of nodes going from 24000 to almost 30000.

Since in this case the information about the links that will appear is available, the evaluation framework differs from the one previously presented. For every snapshot at times $i = [1, 5]$ the algorithms have been executed in order to assign likelihood scores to the non-observed links and the link-prediction performance has been evaluated with respect to every future time point $j = [i+1, 6]$. Considering a pair of time points $(i, j)$, the non-observed links at time $i$ are ranked by decreasing likelihood scores and the precision is computed as the percentage of links that appear at time $j$ among the top-$r$ in the ranking, where $r$ is the total number of non-observed links at time $i$ that appear at time $j$. Non-observed links at time $i$ involving nodes that disappear at time *j* are not considered in the ranking. Table 9 reports for each method a 5-dimensional upper triangular matrix, containing as element (*i, j*) the precision of the link-prediction from time *i* to time *j*+1.

As a confirmation of the results seen on the artificial networks, CH outperformed SPM and ranked first in the prediction for all the pairwise time points, with a mean precision of 0.13 versus 0.09. It can be noticed that the precision improves as the two time points become further, going from 0.11 to 0.14 for CH and from 0.07 to 0.11 for SPM. In addition, on the contrary to what reported for small-size networks where SPM was slightly faster than CH, here the execution time is much smaller for the local method, with a difference of around 5 hours, suggesting that the computational requirements of the global method considerably increase with the network size. This is actually in agreement with the computational complexity of the two methods, since SPM executes in $O(N^3)$ whereas CH, on sparse networks as the ones considered (Table 8 shows the low density), requires only $O(N^2)$ (for further details please refer to the Methods section).

Suppl. Table 1 reports the comparison between CH and RA (Resource-Allocation) [36], which is the respective non-LCP variant and is one of the best parameter-free local methods without considering the LCP-based predictors (see Methods for details). Since from the previous literature it is known that CH generally performs better than RA for small networks [16], [18]–[20], we decided to test their performance also on big networks in order to verify that this is valid also with the increase of the size. The results confirm that the trend still holds, with CH surpassing RA for all the pairwise time points.

To conclude, we want to underline that, differently from the removal and re-prediction framework in which the set of missing links is artificially generated by a random procedure, here the set of links that will appear between two consecutive time points is given by ground-truth information, which makes the result even more significant and truthful, confirming the effectiveness of CH.

## 2.6 Evaluation on large-size real complex networks

Since the above considered Internet networks were characterized by a high number of nodes and the local CH model outperformed the best global model SPM, we advanced the hypothesis that network size could play an important role. Consequently, with a substantial computational effort, we created the first study ever conducted to also perform and include removal and re-prediction evaluation for the best global and local methods on 7 large-size networks (from 3000 up to 40000 nodes), several statistics are shown in Table 8.

Considering the same evaluation framework described for the small-size real networks, Table 10 shows the precision for each network, the mean precision and mean ranking for a further comparison of the two overall performance scores in discussion, and, analogously to Table 4, the p-value of the permutation test (10000 iterations) for the mean ranking.

Looking at the best method (highlighted in bold) for each network, it is evident that CH surpasses SPM in 5 out of 7 networks. This result is confirmed by the mean ranking, 1.29 for CH against 1.71 for SPM, and the p-value of the permutation test supports the significant difference in performance.

On the other side, the mean precision shows again a misleading view and could bring to the opposite conclusion. The main reason is the network *arxiv astroph*, one of the two in which SPM is better. Here indeed, the bigger gap of performance between the two methods nullifies the numerous victories of CH in the other networks. However, the ability of a method to obtain higher performances in multiple networks is an indicator of great robustness and adaptability of the approach to diverse topologies. We believe that these principles should obtain a higher consideration with respect to a method that offers a lower performance on many networks and rare peaks of outperformance in a few networks, which may be even due to overfitting toward certain structural features. These results stress the importance of the introduction of the precision-ranking evaluation framework, as well as representing a further scenario in which CH, the best local approach, overcomes the best global approach, SPM.

Table 11 reports the computational time required by the methods in order to perform the link-prediction. The algorithms have been run in the same workstation, for additional details please refer to the Hardware and software section. The table is a confirmation of the lower computational complexity of CH, as already discussed in the previous section. Noteworthy is the increase of time from *thesaurus* (around 24000 nodes) to *facebook* (around 44000 nodes), where CH goes from 1.3 to 2.1 hours, whereas SPM passes from 2.5 to 15.3 hours, underlining in a tangible way the stronger computational time dependency on the network size.

Suppl. Table 2 reports the comparison between CH and RA, always limited to networks of small size in previous studies and here finally offered also for large size networks. The overall performance is significantly higher for CH, further confirming the current belief supported by many studies [3], [8], [14], [16], [18]–[20] that CH seems the best local parameter-free deterministic method.

# 3. Discussion

Link-prediction studies are not always convincing in the selection of the approaches to adopt as a reference for comparison. Global methods are believed to be the best performing and SBM, despite its remarkably high computational time, is often considered among the best state-of-the-art methods to use as a baseline. However, we could not find in the scientific literature well-grounded proofs of significant outperformance with respect to other state-of-the-art methods. Consequently, we decided to conduct an accurate study that compares SBM with SPM and CH, the two methods that recently were pointed in many studies as the best respectively for global and local link-prediction. In addition, for completeness, a third method named FBM was considered, representing a faster variant of SBM.

In contrast to the malpractice of testing the methods in a reduced benchmark of small-size networks, we presented an extensive analysis evaluating the methods on four different frameworks: re-prediction of randomly removed links both on small-size real networks, small-size artificial networks and large-size real networks, as well as link-prediction of a time-evolving large-size real network.

From the wide investigation several key messages emerged, which will be now summarized. First, SPM demonstrated to be the best global method, significantly outperforming SBM in both real and artificial networks. Second, SBM, commonly adopted as a state-of-the-art baseline for comparison, displayed a huge computational time with respect to the other approaches, without an overall significant gain in prediction performance even versus the best local method CH. Third, the mean precision resulted to be an inappropriate metric of overall performance. In fact, the mean is a central measure affected by the presence of peculiar networks that strongly favour the prediction only for some methods, whereas the precision-ranking provides a more robust and unbiased overview. Fourth, the evaluation on multiple frameworks highlighted that the adoption of a single benchmark with only small-size networks, although the number of networks tested is large, can easily bring to misleading conclusions showing only part of the truth. Last but not least, CH, the best local approach, outperformed the best global approach, SPM, in three out of four evaluation frameworks.

Now, due to the impressive results obtained by a simple yet powerful local, parameter-free and model-based deterministic rule such as CH, we are going to dedicate a paragraph for a thoughtful analysis of the fundamentals that advocate the bioinspired theory upon which is based. In 1949, Donald Olding Hebb advanced a *local-learning-rule* in neuronal networks that can be summarized in the following: neurons that fire together wire together [37]. In practice, the Hebbian learning theory assumes that different engrams (memory traces) are memorized by the differing neurons' cohorts that are co-activated within a given network. Yet, the concept of wiring together was not further specified, and could be interpreted in two different ways. The first interpretation is that the connectivity already present, between neurons that fire together, is reinforced; whereas, the second interpretation is the emergence and formation of new connectivity between non-interacting neurons already embedded in a interacting cohort.

The first interpretation has been demonstrated in several neuroscientific studies, where it was proven that certain forms of learning consist of synaptic modifications, while the number of neurons remains basically unaltered [38]–[40]. A first mathematical model of this learning process was implemented in the Hopfield's model of associative memory, where neuron-assemblies are shaped during engram formation by a re-tuning of the strengths of all the adjacent connections in the network [41]. It is important to specify that neuronal networks are over-simplified models and between two nodes (that represent two neurons) only one unique connection, which is deceptively called 'synapse', is allowed. This unique artificial synapse is a network link with a weight (or strength) and abstractly represents in a unique connectivity all the multitude of synapses that can occur between two real neurons in a brain tissue. For non-computational readers, we stress that the word 'synapse' used in computational modelling of artificial neural networks might be misleading for neurobiologists, and should be intended as a mere link between two nodes of a network that comprehensively symbolizes the strength of all the real biological synapses connecting two neurons. Here, and in the remainder of this article, we will refer only to this artificial neural network model where a link between two nodes (neurons) indicates an abstract interaction between them. In fact, although this artificial network model is based on evident simplifications, it demonstrated to be a powerful tool to simulate learning processes of intelligent systems [41], [42].

Surprisingly, the second possible interpretation of the Hebbian learning – a cohort of interacting neurons that fire together give rise to new connections between non-interacting neurons in the cohort - to the best of our knowledge was never formalized as a general paradigm of learning, and therefore, it was never employed with success to modify the architecture of abstract neural networks to simulate *pure topological learning*. We acknowledge the existence of studies that investigate how neuronal morphology predicts connectivity [43]. For instance, Peters' rule predicts connectivity among neuron types based on the anatomical colocation of their axonal and dendritic arbors, providing a statistical summary of neural circuitry at mesoscopic resolution [43]. However, no paradigms were proposed to explain the extent to which new connections between non-interacting neurons could be predicted in function of their *likelihood* to be collectively co-activated (by firing together) on the already existing network architecture. This likelihood of localized functional interactions on the existing neural network can be influenced by external factors such as the temporal co-occurrence of the firing activity on a certain cohort of neurons, and by other factors that are intrinsic to the network architecture such as, among the most important, the *network topology*.

In 2013, Cannistraci et al. noticed that considering only the network topology, the second interpretation of the Hebbian learning could be formalized as a mere problem of topological link-prediction in complex networks. The rationale is the following. The network topology plays a crucial role in isolating cohorts of neurons in functional communities that naturally and preferentially - by virtue of this predetermined local-community topological organization - can perform local processing. In practice, the local-community organization of the network topology creates a physical and structural 'energy barrier' that allows the neurons to preferentially fire together within a certain community and therefore to add links inside that community, implementing a type of local topological learning. In few words: the local-community organization influences (by increasing) the likelihood that a cohort of neurons fires together because they are confined in the same local-community, and, consequently, also the likelihood that they will create new connections inside the community is increased by the mere structure of the network topology. Inspired by this intuition, Cannistraci et al. called this local topological learning theory *epitopological learning*, which stems from the second interpretation of the Hebbian leaning. The definition was not clearly given in the first article [3] that was immature, and therefore, we now provide an elucidation of the concepts behind this theory by offering new definitions. *Epitopological learning* occurs when cohorts of neurons tend to be preferentially co-activated because they are topologically restricted in a local-community, and therefore, they tend to facilitate learning of new network features by forming new connections instead of merely re-tuning the weights of existing connections. As a key intuition, Cannistraci et al. postulated also that the identification of this form of learning in neuronal networks was only a special case; hence, the *epitopological learning* and the associated *local-community-paradigm (LCP)* were proposed as local rules of learning, organization and link-growth valid in general for topological link-prediction in any complex network with LCP architecture [3]. In the present article, we expand forward this theory proposing that in reality the LCP-based and CN-based rules for link-prediction are much more than methods for link-prediction, but they belong to the general class of local-ring network automata for link-prediction in complex networks that we defined in the first result section.

A particular doubt emerges due to fact that CH is able to outperform SPM on small-size artificial networks and large-size real networks. However, paradoxically, in small-size real networks the contrary is true and SPM outperforms CH. Since the small-size artificial networks were generated using the PSO model, we thought to investigate the hyperbolicity also of the real small-size networks considered in the first evaluation. It is known that the PSO model generates artificial networks with an underlying hyperbolic geometry [32] and large-size real networks, in order to make efficient the navigation and the global information delivery [44], are often characterized by a marked hyperbolic geometry [32], [45]. A representative case is the one of the Internet AS topologies: many studies demonstrated that they have a distinct hyperbolic geometry, in fact the greedy routing efficiency, robustness and scalability is generally maximized when the space is hyperbolic, both in single-layer [45]–[47] and in multiplex networks [48]. However, this might not be necessarily true for the small-size real networks, where, due to the reduced size, the density is high, and therefore their topology cannot often respect a hyperbolic geometry. This is confirmed by the results in Table 1 and Table 8, where it is shown that all the small-size networks have a high density in comparison to large-size networks. Since a peculiar and necessary feature of networks with underlying hyperbolic geometry is a scale-free degree distribution [32], [45], [49] we performed a comparison between the estimated power-law degree distribution exponents of small-size and large-size real networks. As highlighted in Fig. 5, the large-size real networks have a significantly lower exponent (p-value < 0.01) and therefore are characterized by a significantly higher power-lawness than small networks. Furthermore, the large-size networks average value is 2.6, which perfectly falls in the typical range $2 < \gamma < 3$ [33]. On the contrary, small-size networks have a mean exponent of 4.1, which represents an outlier with respect to that range. In brief, CH is a physical rule that exhibits a stronger performance in comparison to a general learning-algorithm in networks characterized by an underlying hyperbolic geometry, hence we speculate that the physical model behind CH might be able to well capture the dynamics of organization of systems with this intrinsic characteristic. In fact, CH might be one the basic principles and generative mechanisms that contributes to give origin to the growth of hyperbolic networks by facilitating the transition from local-tunnels to local-rings and, in turn, generating local-community link-clustering in the network topology. Previous studies demonstrated how bioinspired modelling can capture the basic dynamics of network adaptability through iteration of local rules, and produces in few hours of computing solutions with properties comparable to or better than those of real-world infrastructure networks, which would require many months of designing by teams of engineers [50]. Similarly, this article aims to promote interest for both bioinspired computing and network automata, demonstrating that a simple unsupervised rule that emulates principles of network self-organization and adaptiveness arising during learning in living intelligent systems (like the brain), can equiperform, and sometimes outperform, advanced learning-machines (algorithms based on inference such as SPM, SBM and FBM) that exploit global network information. Furthermore, in support to the more accurate predictions of CH on the time-evolving AS topologies, a recent study highlighted similar optimization principles between synaptic plasticity rules that regulate neural network activity and algorithms commonly used for controlling the flow of data in engineered networks such as Internet [51]. In particular, the additive increase and multiplicative decrease rule (AIMD), which is the congestion control algorithm adopted in the Internet transmission control protocol (TCP) [52], has also strong theoretical and experimental support for long term potentiation (LTP) and long term depression (LTD) in brain [51]. Moreover, the algorithm is very similar to an edge-weight update rule shown to produce stable Hebbian learning compared to many other rules [53], [54]. This similarity was at the moment proven only for changing weights of existing connectivity, hence it represents a *geometrical learning*. Our results are promising because they pave the way to extend the similarity between neural networks and Internet networks architectures also from the mere topological point of view, where, according to the LCP theory and the related epitopological learning, the process of *structural learning* is given by addition or deletion of connectivity.

The results in Fig. 3, Fig. 4 and Suppl. Fig.3, as already discussed, show that in most of the cases CH is able to predict links in the hyperbolic artificial networks (PSO and nPSO) with a precision even higher than the original hyperbolic distances (HD) in presence of communities, and with an almost comparable precision for low temperatures (high clustering) in absence of communities. In our opinion, this is another evidence in support of our hypothesis that CH might be one of the basic principles and generative mechanisms that contributes to give origin to the growth of hyperbolic networks by facilitating the transition from local-tunnels to local-rings and, in turn, growing local-community link-clustering in the network topology. However, science is moved by intuitions, speculations, hypotheses and conjectures that need to be proved. Therefore, here, before closing this article, we want to give a concrete proof based on simulations that CH is a generative rule particularly valid for hyperbolic geometry.

To this aim, we notice that scale-freeness seems a necessary condition for hyperbolicity [45], [49]. This means that non-scale-free networks are non-hyperbolic, therefore theoretically if it is true that CH is a generative rule particularly valid for hyperbolic geometry, then on non-hyperbolic networks the link-prediction performance of CH should be reduced and inferior to SPM, like we noticed in real small-size networks that having a high power-law exponent are weakly hyperbolic. Actually, to be more precise, since real small-size networks seem weakly hyperbolic, CH performance was in general lower than SPM but not 'significantly' lower - from a statistical point of view - because it is very rare to detect real networks that are not scale-free at all and therefore not hyperbolic. However, using an artificial random model of non-scale-free networks - like the Watts-Strogatz model [55] - whose input parameters give the possibility to tune the clustering coefficient - we should be able to prove that the performance of CH is significantly lower than SPM for high levels of clustering. In fact, according to a recent study of Krioukov [56], non-scale-free networks with strong clustering have a latent network geometry that is Euclidean. On the other hand, using the same Watts-Strogatz model with low level of clustering, the random networks lose any latent geometry, and therefore both link predictors should dramatically lose their prediction power in general. This would ultimately demonstrate that latent geometry is at the basis of link-prediction and that SPM performs well for Euclidean latent geometry given by non-scale-free and non-hyperbolic networks, whereas CH performs well for hyperbolic latent geometry given by scale-free hyperbolic networks. Since many real world networks tend to exhibit hyperbolicity, and therefore scale-freeness, this possible simulation on the Watts-Strogatz model would be the final demonstration that the finding that CH seems to perform better than SPM, and global methods, on real networks is true in general and has theoretical well-grounded basis in the latent geometry of the real networks.

We performed the simulation using the Watts-Strogatz model [55] considering parameters $N$ = [100, 500, 1000] (network size), $m$ = [10, 12, 14] (half of average degree) and $\beta$ = [0.001, 0.01, 0.1] (rewiring probability). The values chosen for the parameters $N$ and $m$ are the same used for the synthetic networks generated using the PSO model. The values chosen for $\beta$ are intended to produce networks with different properties mainly in terms of clustering coefficient and characteristic path length. Fig. 6 reports the link-prediction results and it is evident that the simulation using the artificial model solidly proves what we theoretically prefigured in the paragraph above.

Suppl. Fig. 4 shows that also in this framework SPM and CH obtain overall the best and more robust performance with respect to SBM, which has a particular drop in precision for N = 500-1000, and FBM, whose discrepancy with respect to the other methods is huge for N = 100. To notice that the SBM-FBM trend is the same obtained for the link-prediction on the PSO model, where SBM is better than FBM for N = 100 and their ranking is inverted for N = 500-1000.

This last test confirms that SBM, regardless of the generative model used for creating the artificial networks, does not perform at the same level as the other methods, and therefore as a model-learning machine it is not able to generalize enough, at least not as much as SPM for link-prediction. This suggests that SBM overfits the structure of the network used for learning. In fact, a recent study [57] highlighted that if only the single partition with the highest posterior probability is used for predicting the links, the performance decreases with respect to the case in which an ensemble of likely partitions are considered (as in the algorithm here adopted), because the single partition significantly overfits the network. What our simulations additionally spot out is that, although the ensemble procedure should mitigate the overfitting, it seems that it still remains a major drawback of the algorithm.

Coming to the conclusions, on the basis of the results emerged from this extensive analysis, we would like to suggest some guidelines to follow in new coming link-prediction studies. First, we invite to reject the widespread consideration of SBM as a state-of-the-art baseline for a performance comparison with new proposed methods. Our wide evaluations highly support the significant outperformance of SPM and CH respectively as best global and local methods, therefore we strongly encourage their adoption as references for a fair comparison to the state-of-the-art. Second, in order to prevent erroneous conclusions, we stress the importance to follow a robust evaluation framework, based on multiple types of link-prediction evaluations. Methods should be tested on different fronts, for instance re-prediction of randomly removed links and prediction in time-evolving networks. Both real networks and artificial models should be taken into account, considering a rich benchmark dataset that ranges over different network sizes and diverse topological features. Among the ones tested, the nonuniform PSO model turned out to be the closest to generate artificial networks with realistic topologies. Precision should be adopted as metric of evaluation on the single network and, while comparing the methods over several networks, the precision-ranking should be used to obtain an unbiased score of overall performance.

Finally, we close the discussion with an open question. The detailed analysis of the links correctly predicted by the methods suggested that on average only half of them overlap between two different approaches, whereas the remaining part is peculiar of a single method and distributed in a similar abundance among the two. This offers a margin of improvement that could be exploited by a proper combination of methods and paves the way for the investigation of hybrid approaches potentially able to reach higher performances in topological link-prediction. But, we seriously believe that to build these 'intelligent hybrid methods' for topological link-prediction, we urge to found the basis of a latent geometry theory of link-prediction in complex networks, and this study is aimed to be a first landmark that points towards this direction.

## 4. Methods

### 4.1 Cannistraci-Resource-Allocation (CRA) and Cannistraci-Hebb (CH) network automata model

Cannistraci-Resource-Allocation (CRA) is a local-based, parameter-free and model-based deterministic rule for topological link-prediction in both monopartite [3] and bipartite networks [4], [5]. It is based on the *local-community-paradigm* (LCP) which is a bioinspired theory recently proposed in order to model local-topology-dependent link-growth in a class of real complex networks characterized by the development of diverse, overlapping and hierarchically organized local-communities [3]. Being a local-community-based method, it assigns to every candidate interaction a likelihood score looking only at the neighbours nodes and their cross-interactions. In particular, the paradigmatic shift introduced by the LCP is to consider not only the common-neighbours of the interacting nodes but also the links between those common-neighbours, which in practice form all together a local-community (Fig. 2). In this article, we explained that the rule behind CRA can be more in generally interpreted as the generative rule of a network automaton that we named Cannistraci-Hebb (CH). And the local-community coincides with the central portion of the local-tunnel that consists only of the CNs and LCLs. For each candidate interaction between nodes $i$ and $j$, the score is assigned according to the following equation [3]:

$$CH(i,j) = \sum_{k \in \Phi(i) \cap \Phi(j)} \frac{|\varphi(k)|}{|\Phi(k)|} \qquad (1)$$

Where:

$k$: common neighbour of nodes $i$ and $j$

$\Phi(i)$: set of neighbours of node $i$

$|\Phi(k)|$: cardinality of set $\Phi(k)$, equivalent to the degree of $k$

$\varphi(k)$: sub-set of neighbours of $k$ that are also common-neighbours of $i$ and $j$

$|\varphi(k)|$: equivalent to the local-community degree of $k$ (see Fig. 2)

The higher the CH score, the higher the likelihood that the interaction exists, therefore the candidate interactions are ranked by decreasing CH scores and the obtained ranking is the link-prediction result.

The computational complexity of the CH method is *O(EN\*(1-D))*, where *N* and *E* are the number of nodes and links in the network, and $D = \frac{2E}{N(N-1)}$ is the network density. However, in the domain of real and practical problems in which topological link-prediction is applied, the complexity of CH can be more simply expressed as $O(EN)$, and frequently approximated to $O(N^2)$, for further details please refer to the Appendix A.

Note that the link likelihoods are computed independently from each other and therefore the implementation can be easily parallelized in order to speed up the running time. The method has been implemented in MATLAB. The code is available at: https://sites.google.com/site/carlovittoriocannistraci/

## 4.2 Structural Perturbation Method (SPM)

SPM is a structural perturbation method that relies on a theory similar to the first-order perturbation in quantum mechanics [9]. Unlike CH, it is a global approach, meaning that it exploits the information of the complete adjacency matrix in order to compute the likelihood score to assign to every candidate interaction. A high-level description of the procedure is the following:

1) Randomly remove a subset of the edges $\Delta E$ (usually 10%) from the network adjacency matrix $x$, obtaining a reduced adjacency matrix $x^R$.
2) Compute the eigenvalues and eigenvectors of $x^R$.
3) Considering $\Delta E$ as a perturbation of $x^R$, construct the perturbed matrix $\tilde{x}$ via a first-order approximation that allows the eigenvalues to change while keeping fixed the eigenvectors.
4) Repeat steps 1-3 for 10 independent iterations and take the average of the perturbed matrices $\tilde{x}$.

The idea behind the method is that a missing part of the network is predictable if it does not significantly change the structural features of the observable part, represented by the eigenvectors of the matrix. If this is the case, the perturbed matrices should be good approximations of the original network [9]. The entries of the average perturbed matrix represents the scores for the candidate links. The higher the score the greater the likelihood that the interaction exists, therefore the candidate interactions are ranked by decreasing scores and the obtained ranking represents the link-prediction result.

The computational complexity of the SPM method is $O(kN^3)$, where $k$ is the number of iterations and $N$ is the number of nodes in the network. In fact, every iteration is dominated the eigen-decomposition of the perturbed adjacency matrix, which requires $O(N^3)$. However, $k$ is usually a small constant (e.g. 10) and the iterations, since independent from each other, can be executed in parallel in order to speed up the running time.

The MATLAB implementation of the method has been provided by the authors.

## 4.3 Stochastic Block Model (SBM)

The framework based on stochastic block model (SBM) considered in this study has been introduced by Guimerà et al. [7] in order to identify both missing and spurious interactions in noisy network observations. The general idea of a block model is that the nodes are partitioned into groups and the probability that two nodes are connected depends only on the groups to which they belong. Assuming that there is no prior knowledge about which partition is more suitable for the observed network, the mathematical formula for obtaining the reliability of an individual link between nodes $i$ and $j$ is [7]:

$$R_{ij} = \frac{1}{Z} \sum_{p \in P} \left( \frac{l_{\sigma_i \sigma_j} + 1}{r_{\sigma_i \sigma_j} + 2} \right) exp[-H(p)] \qquad (2)$$

Where the sum is over every partition $p$ in the space $P$ of all the possible partitions of the network into groups, $\sigma_i$ is the group of node $i$ in partition $p$, $l_{\alpha\beta}$ is the number of links between groups $\alpha$ and $\beta$, $r_{\alpha\beta}$ is the maximum number of possible links between groups $\alpha$ and $\beta$. The function $H(p)$ is:

$$H(p) = \sum_{\alpha \leq \beta} \left[ ln(r_{\alpha\beta} + 1) + ln \binom{r_{\alpha\beta}}{l_{\alpha\beta}} \right] \qquad (3)$$

And the normalization factor is:

$$Z = \sum_{p \in P} exp[-H(p)] \qquad (4)$$

However, since the exploration of all the possible partitions is too computationally expensive, the Metropolis algorithm, which is based on a stochastic procedure, is exploited in order to sample only a subset of partitions that are relevant for the estimation of the link reliability [7]. The higher the reliability the greater the likelihood that a non-observed interaction actually exists, therefore the candidate interactions are ranked by decreasing scores and the obtained ranking represents the link-prediction result. The C code of the method has been released by the authors and can be download from the website http://seeslab.info/downloads/network-c-libraries-rgraph/.

## 4.4 Fast probability Block Model (FBM)

Fast probability Block Model (FBM) is a global method based on the same network partitioning theory as SBM, but it replaces the Metropolis algorithm introducing a greedy strategy for an efficient sampling over the space of the possible partitions, which leads to high improvements in the computational time [8].

For each network 50 partitions are sampled according to the following procedure. As first the network is randomly partitioned in two blocks. Then, for each block, until all its edges have been considered, the maximum clique is iteratively removed and it represents a group for the current partitioning. At the end of the iterative removal a set of low degree nodes will remain without forming any clique, they are treated as a separate special group having low inner link density [8].

Given the sampled partitions, the following mathematical formula is used in order to compute the likelihood of the non-observed links [8]:

$$R_{ij} = \frac{1}{|P|} \sum_{p \in P} F(\sigma_i, \sigma_j) \tag{5}$$

$$F(\alpha, \beta) = \begin{cases} \dfrac{r_\alpha}{2r_\alpha - l_\alpha}, & \alpha = \beta \\ \dfrac{l}{r_{\alpha\beta} - l_{\alpha\beta}}, & \alpha \neq \beta \end{cases} \tag{6}$$

Where the sum is over every partition $p$ in the set $P$ of sampled partitions, $\sigma_i$ is the group of node $i$ in partition $p$, $l_\alpha$ is the number of links within group $\alpha$, $r_\alpha$ is the maximum number of possible links within group $\alpha$, $l_{\alpha\beta}$ is the number of links between groups $\alpha$ and $\beta$, $r_{\alpha\beta}$ is the maximum number of possible links between groups $\alpha$ and $\beta$.

The higher the reliability the greater the likelihood that a non-observed interaction actually exists, therefore the candidate interactions are ranked by decreasing scores and the obtained ranking represents the link-prediction result. The MATLAB implementation of the method has been provided by the authors.

### 4.5 Resource-Allocation (RA)

Resource-Allocation (RA) is a local-based, parameter-free and model-based deterministic rule for topological link-prediction [36], motivated by the resource allocation process taking place in networks. Considering a pair of nodes that are not directly connected, one node can send some resource to the other one through their common-neighbours, which play the role of transmitters. It assumes the simplest case where every transmitter equally distributes a unit of resource between its neighbours. For each candidate interaction between nodes $i$ and $j$, the score is assigned according to the following equation [36]:

$$RA(i,j) = \sum_{k \in \Phi(i) \cap \Phi(j)} \frac{1}{|\Phi(k)|} \tag{7}$$

Where:

$k$: common neighbour of nodes $i$ and $j$

$\Phi(i)$: set of neighbours of node $i$

$|\Phi(k)|$: cardinality of set $\Phi(k)$, equivalent to the degree of $k$

The higher the RA score, the higher the likelihood that the interaction exists, therefore the candidate interactions are ranked by decreasing RA scores and the obtained ranking is the link-prediction result. The method has been implemented in MATLAB.

### 4.6. Generation of synthetic networks using the PSO model

The Popularity-Similarity-Optimization (PSO) model [32]p is a generative network model recently introduced in order to describe how random geometric graphs grow in the hyperbolic space. In this model the networks evolve optimizing a trade-off between node popularity, abstracted by the radial coordinate, and similarity, represented by the angular coordinate distance, and they exhibit many common structural and dynamical characteristics of real networks.

The model has four input parameters:

- $m > 0$, which is equal to half of the average node degree;
- $\beta \in (0, 1]$, defining the exponent $\gamma = 1 + 1/\beta$ of the power-law degree distribution;
- $T \geq 0$, which controls the network clustering; the network clustering is maximized at $T = 0$, it decreases almost linearly for $T = [0,1)$ and it becomes asymptotically zero if $T > 1$;
- $\zeta = \sqrt{-K} > 0$, where $K$ is the curvature of the hyperbolic plane. Since changing $\zeta$ rescales the node radial coordinates and this does not affect the topological properties of networks [32], we considered $K = -1$.

Building a network of $N$ nodes on the hyperbolic disk requires the following steps:

(1) Initially the network is empty;

(2) At time $i = 1, 2, \dots, N$ a new node $i$ appears with radial coordinate $r_i = 2ln(i)$ and angular coordinate $\theta_i$ uniformly sampled in $[0,2\pi]$; all the existing nodes $j < i$ increase their radial coordinates according to $r_j(i) = \beta r_j + (1 - \beta)r_i$ in order to simulate popularity fading;

(3) If $T = 0$, the new node connects to the $m$ hyperbolically closest nodes; if $T > 0$, the new node picks a randomly chosen existing node $j < i$ and, given that it is not already connected to it, it connects to it with probability

$$p(i,j) = \frac{1}{1 + \exp\left(\frac{h_{ij} - R_i}{2T}\right)} \tag{8}$$

repeating the procedure until it becomes connected to *m* nodes.

Note that

$$R_i = r_i - 2\ln\left[\frac{2T(1 - e^{-(1-\beta)\ln(i)})}{\sin(T\pi)\,m(1-\beta)}\right] \tag{9}$$

is the current radius of the hyperbolic disk, and

$$h_{ij} = arccosh(\cosh r_i \cosh r_j - \sinh r_i \sinh r_j \cos \theta_{ij}) \tag{10}$$

is the hyperbolic distance between node *i* and node *j*, where

$$\theta_{ij} = \pi - \left|\pi - \left|\theta_i - \theta_j\right|\right| \tag{11}$$

is the angle between these nodes.

(4) The growing process stops when N nodes have been introduced.

### 4.7. Generation of synthetic networks using the nonuniform PSO (nPSO) model

The nonuniform PSO (nPSO) model [34] is a variation of the PSO model introduced in order to confer to the generated networks an adequate community structure, which is lacking in the original model. Since the connection probabilities are inversely proportional to the hyperbolic distances, a uniform distribution of the nodes over the hyperbolic disk does not create agglomerates of nodes that are concentrated on angular sectors and that are more densely connected between each other than with the rest of the network. A nonuniform distribution, instead, allows to do it by generating heterogeneity in angular node arrangement. In particular, without loss of generality, we will concentrate on the Gaussian mixture distribution, which we consider suitable for describing how to build a nonuniform distributed sample of nodes along the angular coordinates of a hyperbolic disk, with communities that emerge in correspondence of the different Gaussians.

A Gaussian mixture distribution is characterized by the following parameters [58]:

- $C > 0$, which is the number of components, each one representative of a community;
- $\mu_{1\ldots C} \in [0, 2\pi]$, which are the means of every component, representing the central locations of the communities in the angular space;
- $\sigma_{1\ldots C} > 0$, which are the standard deviations of every component, determining how much the communities are spread in the angular space; a low value leads to isolated communities, a high value makes the adjacent communities to overlap;

- $\rho_{1...C}$ ($\sum_i \rho_i = 1$), which are the mixing proportions of every component, determining the relative sizes of the communities.

Given the parameters of the PSO model $(m, \beta, T)$ and the parameters of the Gaussian mixture distribution $(C, \mu_{1...C}, \sigma_{1...C}, \rho_{1...C})$, the procedure to generate a network of $N$ nodes is the same described in the section for the uniform case, with the only difference that the angular coordinates of the nodes are not sampled uniformly but according to the Gaussian mixture distribution. Note that, although the means of the components are located in $[0, 2\pi]$, the sampling of the angular coordinate $\theta$ can fall out of this range. In this case, it has to be shifted within the original range, as follows:

- If $\theta < 0 \rightarrow \theta = 2\pi - mod(|\theta|, 2\pi)$
- If $\theta > 2\pi \rightarrow \theta = mod(\theta, 2\pi)$

Although the parameters of the Gaussian mixture distribution allow for the investigation of disparate scenarios, as a first case of study we focused on the most straightforward setting. For a given number of components $C$, we considered their means equidistantly arranged over the angular space, the same standard deviation and equal mixing proportions:

- $\mu_i = \frac{2\pi}{C} * (i - 1) \quad i = 1 ... C$
- $\sigma_1 = \sigma_2 = ... = \sigma_C = \sigma$
- $\rho_1 = \rho_2 = ... = \rho_C = \frac{1}{C}$

In particular, in our simulations we fixed the standard deviation to 1/6 of the distance between two adjacent means $\left(\sigma = \frac{1}{6} * \frac{2\pi}{C}\right)$, which allowed for a reasonable isolation of the communities. The community memberships are assigned considering for each node the component whose mean is the closest in the angular space.

### 4.8 Generation of synthetic networks using the Watts-Strogatz model

The Watts-Strogatz model [55], proposed in 1998, introduced the concept of small-world networks, strongly clustered as regular lattices and with a small characteristic path length like random graphs, arguing that many real networks are somewhere between these two extreme topological configurations.

It has three input parameters: $N$, which is the number of nodes; $m > 0$, representing half of the average node degree and therefore defining the number of edges $E = mN$; $\beta \in [0, 1]$, which is the rewiring probability.

The procedure to generate a network requires the following steps: (1) Create a ring lattice of $N$ nodes, assuming the nodes ordered in a circular list and connecting each of them to its $m$ next and previous neighbours; (2) For every node, consider each edge to the $m$ next neighbours and rewire it with probability $\beta$. The new target node is chosen uniformly at random, avoiding self-loops and link duplication.

Tuning the parameter $\beta$ allows to generate networks with characteristics between regularity ($\beta = 0$, no links rewired) and randomness ($\beta = 1$, all the links rewired). In particular, Watts and Strogatz [55] showed how, starting from a ring lattice, the introduction of even a few (small $\beta$) short-cuts leads to an immediate drop in the characteristic path length, whereas the high clustering coefficient remains practically unchanged. They are random networks with not-scale-free node distribution.

### 4.9 Real networks datasets

The real networks have been transformed into undirected and unweighted, self-loops have been removed and the largest connected component has been considered.

*Mouse neural*: in-vivo single neuron connectome that reports mouse primary visual cortex (layers 1, 2/3 and upper 4) synaptic connections between neurons [59].

*Karate*: social network of a university karate club collected by Wayne Zachary in 1977. Each node represents a member of the club and each edge represents a tie between two members of the club [60].

*Dolphins*: a social network of bottlenose dolphins. The nodes are the bottlenose dolphins (genus Tursiops) of a bottlenose dolphin community living off Doubtful Sound, a fjord in New Zealand. An edge indicates a frequent association. The dolphins were observed between 1994 and 2001 [61].

*Macaque neural*: a macaque cortical connectome, assembled in previous studies in order to merge partial information obtained from disparate literature and database sources [62].

*Polbooks*: nodes represent books about US politics sold by the online bookseller Amazon.com. Edges represent frequent co-purchasing of books by the same buyers, as indicated by the "customers who bought this book also bought these other books" feature on Amazon. The network was compiled by V. Krebs and is unpublished, but can found at http://www-personal.umich.edu/~mejn/netdata/.

*ACM2009_contacts*: network of face-to-face contacts (active for at least 20 seconds) of the attendees of the ACM Conference on Hypertext and Hypermedia 2009 [63].

*Football*: network of American football games between Division IA colleges during regular season Fall 2000 [64].

*Physicians innovation*: the network captures innovation spread among physicians in the towns in Illinois, Peoria, Bloomington, Quincy and Galesburg. The data was collected in 1966. A node represents a physician and an edge between two physicians shows that the left physician told that the right physician is his friend or that he turns to the right physician if he needs advice or is interested in a discussion [65].

*Manufacturing email*: email communication network between employees of a mid-sized manufacturing company [66].

*Littlerock foodweb*: food web of Little Rock Lake, Wisconsin in the United States of America. Nodes are autotrophs, herbivores, carnivores and decomposers; links represent food sources [67].

*Jazz*: collaboration network between Jazz musicians. Each node is a Jazz musician and an edge denotes that two musicians have played together in a band. The data was collected in 2003 [68].

*Residence hall friends*: friendship network between residents living at a residence hall located on the Australian National University campus [69].

*Haggle contacts*: contacts between people measured by carried wireless devices. A node represents a person and an edge between two persons shows that there was a contact between them [70].

*Worm nervous*: a C. *Elegans* connectome representing synaptic interactions between neurons [55].

*Netsci*: a co-authorship network of scientists working on networks science [71].

*Infectious contacts*: network of face-to-face contacts (active for at least 20 seconds) of people during the exhibition INFECTIOUS: STAY AWAY in 2009 at the Science Gallery in Dublin [63].

*Flightmap*: a network of flights between American and Canadian cities [72].

*Email*: email communication network at the University Rovira i Virgili in Tarragona in the south of Catalonia in Spain. Nodes are users and each edge represents that at least one email was sent [73].

*Polblog*: a network of front-page hyperlinks between blogs in the context of the 2004 US election. A node represents a blog and an edge represents a hyperlink between two blogs [74].

*Odlis*: Online Dictionary of Library and Information Science (ODLIS): ODLIS is designed to be a hypertext reference resource for library and information science professionals, university students and faculty, and users of all types of libraries. Version December 2000 [75].

*Advogato*: a trust network of the online community platform Advogato for developers of free software launched in 1999. Nodes are users of Advogato and the edges represent trust relationships [76].

*Arxiv astroph*: collaboration graph of authors of scientific papers from the arXiv's Astrophysics (astro-ph) section. An edge between two authors represents a common publication [77].

*Thesaurus*: this is the Edinburgh Associative Thesaurus. Nodes are English words, and a directed link from A to B denotes that the word B was given as a response to the stimulus word A in user experiments [78].

*Arxiv hepth*: this is the network of publications in the arXiv's High Energy Physics – Theory (hep-th) section. The links that connect the publications are citations [77].

*Facebook*: a network of a small subset of posts to user's walls on Facebook. The nodes of the network are Facebook users, and each edge represents one post, linking the users writing a post to the users whose wall the post is written on [79].

*ARK200909-ARK201012*: Autonomous systems (AS) Internet topologies extracted from the data collected by the Archipelago active measurement infrastructure (ARK) developed by CAIDA, from September 2009 up to December 2010. The connections in the topology are not physical but logical, representing AS relationships [35].

Most of the networks in the dataset can be downloaded from the Koblenz Network Collection at http://konect.uni-koblenz.de.

**Hardware and software**

Unless stated otherwise, MATLAB code was used for all the simulations. The simulations on small-size networks have been carried out on a Dell workstation under Windows 7 professional 64-bit with 24 GB of RAM and one Intel(R) Xenon(R) X5660 processor with 2.80 GHz. The simulations on large-size networks have been carried out on a workstation under Windows 8.1 Pro with 512 GB of RAM and 2 Intel(R) Xenon(R) CPU E5-2687W v3 processors with 3.10 GHz.

**Funding**

Work in the CVC laboratory was supported by the independent research group leader starting grant of the Technische Universität Dresden. AM was partially supported by the funding provided by the Free State of Saxony in accordance with the Saxon Scholarship Program Regulation, awarded by the Studentenwerk Dresden based on the recommendation of the board of the Graduate Academy of TU Dresden.

**Acknowledgements**

C.V.C. thanks Giancarlo Ferrigno for advancing the question on how the arise of new connections could implement a learning rule in artificial neural networks. C.V.C. thanks Antonio Malgaroli for introducing him to the basic notions of neurobiology of learning. C.V.C. particularly thanks Pierre Julius Magistretti for the precious suggestions and the valuable lectures he personally gave him on neurobiology of learning. C.V.C. thanks Corrado Calì for introducing him to the Peters' rule. C.V.C. thanks Hubert Fumelli for suggesting to clarify the idea of connectivity and synapsis in artificial models. C.V.C. thanks Petra Vertes and Edward Bullmore for suggesting to clarify the idea of LCP and epitopological learning in neuronal networks. C.V.C. thanks Saket Navlakha for the interesting discussions on the difference between LCP theory and clustering. C.V.C. thanks Maksim Kitsak, Anna Korhonen, James Procter and Anton Feenstra for the useful discussion on the definition of CNs in bipartite topology. C.V.C. thanks Timothy Ravasi for encouraging and supporting in the past his research, giving him the freedom to invent and develop the LCP theory. C.V.C. thanks Peter Csermely for his interest and support to the Local-community-paradigm theory. C.V.C. thanks Suzanne Eaton for the inspiring chat on cellular-automata rules.

We are particularly grateful to Fragkiskos Papadopoulos for the AS Internet data and Dmitri Krioukov for the enlightening theoretical discussion on the relation between scale-freeness and hyperbolicity.

We thank Alexander Mestiashvili and the BIOTEC System Administrators for their IT support, Claudia Matthes for the administrative assistance and the Centre for Information Services and High Performance Computing (ZIH) of the TUD.

**Author contributions**

CVC conceived the theory and experiments, planned, directed and supervised the study. CVC conceived and formalized the new idea and definitions behind the local-ring network automata and AM controlled its correctness. AM performed the time complexity analysis and CVC controlled its correctness. AM implemented the code and performed the computational analysis. Both the authors analysed, interpreted the results and contributed in writing the article.

**Competing interests**

The authors declare no competing financial interests.

**Appendix A. CH complexity**

The CH algorithm for topological link-prediction consists of a main loop going over all the non-observed links and at every iteration it independently evaluates the likelihood of one link. Given the number of nodes $N$ and the number of observed links $E$, the number of iterations is:

$$\frac{N(N-1)}{2} - E$$

Considering an iteration in which the non-observed link between two nodes $i$ and $j$ is evaluated, the dominant operation is the intersection between the two sets of neighbours for finding the common-neighbours between $i$ and $j$.

Since the set intersection complexity is linear in the number of elements, the cost is:

$$O(k_i + k_j)$$

Where $k_i$ and $k_j$ are the degrees of the nodes $i$ and $j$.

Although different iterations could have different costs, the average complexity will be:

$$O(2 * avg_k) = O\left(4 * \frac{E}{N}\right) = O\left(\frac{E}{N}\right)$$

Where $avg_k$ is the average node degree.

Given that there are $\frac{N(N-1)}{2} - E$ iterations with average complexity $O\left(\frac{E}{N}\right)$, the overall complexity is:

$$O\left(\left(\frac{N(N-1)}{2} - E\right) * \frac{E}{N}\right) = O\left(\frac{E(N-1)}{2} - \frac{E^2}{N}\right)$$

Gathering the factor $\frac{E(N-1)}{2}$ we obtain:

$$O\left(\frac{E(N-1)}{2}\left(1 - \frac{2E}{N(N-1)}\right)\right) = O\left(\frac{E(N-1)}{2}(1-D)\right)$$

Where D is the network density $D = \frac{2E}{N(N-1)}$.

Removing the multiplicative factor $\frac{1}{2}$ and the constant -1 on which $N$ is dominant, we can rewrite in a more compact form:

$$O(EN(1-D))$$

Let's analyse the complexity in three particular cases:

(1) Minimum number of links for a connected network (tree): $E = N - 1$

$$O\left(\frac{E(N-1)}{2}\left(1 - \frac{2E}{N(N-1)}\right)\right) = O\left(\frac{(N-1)^2}{2}\left(1 - \frac{2(N-1)}{N(N-1)}\right)\right)$$

$$= O\left(\frac{(N-1)^2}{2} - \frac{(N-1)^2}{N}\right) = O\left(N^2 - \frac{N^2}{N}\right) = O(N^2)$$

(2) Half of the number of possible links: $E = \frac{N(N-1)}{4}$

$$O\left(\frac{E(N-1)}{2}\left(1 - \frac{2E}{N(N-1)}\right)\right) = O\left(\frac{N(N-1)^2}{8}\left(1 - \frac{1}{2}\right)\right) = O\left(\frac{N(N-1)^2}{16}\right) = O(N^3)$$

(3) Fully connected network (no non-observed links to evaluate): $E = \frac{N(N-1)}{2}$

$$O\left(\frac{E(N-1)}{2}\left(1 - \frac{2E}{N(N-1)}\right)\right) = O\left(\frac{N(N-1)^2}{8}(1-1)\right) = 0$$

The analysis of the complexity function highlights that the complexity is $O(N^2)$ for sparse networks, it increases as the number of links increases reaching $O(N^3)$ for middle density, and then decreases arriving at a null computational cost at the maximum density, since there are not non-observed links to evaluate.

Due to the fact that reasonable values of density for real-networks are much lower than 0.5, as confirmed by Table 1 and Table 8, we may assert that within the domain of real and practical problems in which topological link-prediction is applied, the complexity of CH can be more simply expressed as $O(EN)$, and very often approximated by $O(N^2)$

Note that the link likelihoods are computed independently from each other and therefore the implementation can be easily parallelized in order to speed up the running time.

**Table 1. Statistics of small-size real networks.**
For each network several statistics have been computed. *N* is the number of nodes. *E* is the number of edges. The parameter *m*, as in the PSO model, refers to half of the average node degree. *D* is the network density. *C* is the average clustering coefficient, computed for each node as the number of links between its neighbours over the number of possible links [55]. *L* is the characteristic path length of the network [55]. *LCP-corr* is the Local-Community-Paradigm correlation [3], representing the correlation between the number of common-neighbours and the number of links between them, looking at each pair of connected nodes in the network. *Struct-cons* is the structural consistency [9], a quantitative index that estimates the link predictability of the network. Power-law is the exponent $\gamma$ of the power-law distribution estimated from the observed degree distribution of the network using the maximum likelihood procedure described in [33].

| | N | E | m | D | C | L | LCP corr | Struct cons | Power law |
|---|---|---|---|---|---|---|---|---|---|
| mouse neural | 18 | 37 | 2.06 | 0.24 | 0.22 | 1.97 | 0.91 | 0.41 | 4.01 |
| karate | 34 | 78 | 2.29 | 0.14 | 0.57 | 2.41 | 0.76 | 0.42 | 2.12 |
| dolphins | 62 | 159 | 2.56 | 0.08 | 0.26 | 3.36 | 0.91 | 0.37 | 6.96 |
| macaque neural | 94 | 1515 | 16.12 | 0.35 | 0.77 | 1.77 | 0.97 | 0.76 | 4.46 |
| polbooks | 105 | 441 | 4.20 | 0.08 | 0.49 | 3.08 | 0.94 | 0.31 | 2.62 |
| ACM2009 contacts | 113 | 2196 | 19.43 | 0.35 | 0.53 | 1.66 | 0.97 | 0.33 | 3.74 |
| football | 115 | 613 | 5.33 | 0.09 | 0.40 | 2.51 | 0.89 | 0.45 | 9.09 |
| physicians innovation | 117 | 465 | 3.97 | 0.07 | 0.22 | 2.59 | 0.79 | 0.21 | 4.51 |
| manufacturing email | 167 | 3250 | 19.46 | 0.23 | 0.59 | 1.97 | 0.99 | 0.55 | 3.13 |
| littlerock foodweb | 183 | 2434 | 13.30 | 0.15 | 0.32 | 2.15 | 0.92 | 0.85 | 3.00 |
| jazz | 198 | 2742 | 13.85 | 0.14 | 0.62 | 2.24 | 0.95 | 0.70 | 4.48 |
| residence hall friends | 217 | 1839 | 8.47 | 0.08 | 0.36 | 2.39 | 0.90 | 0.35 | 6.32 |
| haggle contacts | 274 | 2124 | 7.75 | 0.06 | 0.63 | 2.42 | 0.99 | 0.60 | 1.51 |
| worm nervoussys | 297 | 2148 | 7.23 | 0.05 | 0.29 | 2.46 | 0.91 | 0.23 | 3.34 |
| netsci | 379 | 914 | 2.41 | 0.01 | 0.74 | 6.04 | 0.92 | 0.59 | 3.36 |
| infectious contacts | 410 | 2765 | 6.74 | 0.03 | 0.46 | 3.63 | 0.95 | 0.41 | 6.42 |
| flightmap | 456 | 37947 | 83.22 | 0.37 | 0.81 | 1.64 | 0.99 | 0.78 | 1.71 |
| email | 1133 | 5451 | 4.81 | 0.01 | 0.22 | 3.61 | 0.85 | 0.18 | 4.89 |
| polblog | 1222 | 16714 | 13.68 | 0.02 | 0.32 | 2.74 | 0.93 | 0.25 | 2.38 |

**Table 2. Precision evaluation of link-prediction on small-size real networks.**
For each network 10% of links have been randomly removed (10 iterations for SBM due to the high computational time, 100 iterations for the other methods) and the algorithms have been executed in order to assign likelihood scores to the non-observed links in these reduced networks. In order to evaluate the performance, the links are ranked by likelihood scores and the precision is computed as the percentage of removed links among the top-$r$ in the ranking, where $r$ is the total number of links removed. The table reports for each network the mean precision over the random iterations and the mean precision over the entire dataset. In order to support the discussion in the article, the mean precision excluding *littlerock foodweb* is also shown. For each network the best method (or methods) is highlighted in bold. The networks are sorted by increasing number of nodes $N$.

|  | SPM | SBM | CH | FBM |
|---|---|---|---|---|
| mouse neural | 0.02 | 0.10 | **0.11** | 0.01 |
| karate | 0.17 | **0.28** | 0.20 | 0.27 |
| dolphins | 0.13 | 0.16 | 0.14 | **0.19** |
| macaque neural | **0.72** | 0.68 | 0.56 | 0.55 |
| polbooks | 0.17 | 0.15 | 0.17 | **0.18** |
| ACM2009 contacts | 0.26 | 0.25 | **0.27** | 0.26 |
| football | 0.31 | 0.34 | **0.36** | 0.25 |
| physicians innovation | 0.07 | 0.06 | 0.07 | **0.08** |
| manufacturing email | **0.51** | 0.47 | 0.42 | 0.39 |
| littlerock foodweb | **0.84** | 0.73 | 0.15 | 0.17 |
| jazz | **0.65** | 0.47 | 0.56 | 0.45 |
| residence hall friends | **0.28** | 0.18 | 0.24 | 0.24 |
| haggle contacts | **0.62** | **0.62** | 0.57 | 0.57 |
| worm nervoussys | **0.16** | 0.15 | 0.12 | 0.11 |
| netsci | 0.41 | 0.13 | **0.50** | 0.33 |
| infectious contacts | **0.37** | 0.30 | 0.34 | 0.33 |
| flightmap | **0.75** | 0.64 | 0.54 | 0.56 |
| email | **0.16** | 0.09 | **0.16** | **0.16** |
| polblog | **0.23** | 0.19 | 0.17 | 0.17 |
| mean precision | **0.36** | 0.32 | 0.30 | 0.28 |
| mean precision (without littlerock foodweb) | **0.33** | 0.29 | 0.31 | 0.28 |

**Table 3. Precision-ranking evaluation of link-prediction on small-size real networks.**
For each network 10% of links have been randomly removed (10 iterations for SBM due to the high computational time, 100 iterations for the other methods) and the algorithms have been executed in order to assign likelihood scores to the non-observed links in these reduced networks. In order to evaluate the performance, the links are ranked by likelihood scores and the precision is computed as the percentage of removed links among the top-*r* in the ranking, where *r* is the total number of links removed. The table reports for each network the ranking of the methods by decreasing mean precision (over the random iterations), considering an average rank in case of ties. The mean ranking of the methods over all the networks represents the final evaluation for a proper comparison of the performance. In order to support the discussion in the article, the mean ranking excluding *littlerock foodweb* is also shown. For each network the best method (or methods) is highlighted in bold. The networks are sorted by increasing number of nodes *N*.

|  | SPM | CH | SBM | FBM |
|---|---|---|---|---|
| **mouse neural** | 3 | **1** | 2 | 4 |
| **karate** | 4 | 3 | **1** | 2 |
| **dolphins** | 4 | 3 | 2 | **1** |
| **macaque neural** | **1** | 3 | 2 | 4 |
| **polbooks** | 2.5 | 2.5 | 4 | **1** |
| **ACM2009 contacts** | 2.5 | **1** | 4 | 2.5 |
| **football** | 3 | **1** | 2 | 4 |
| **physicians innovation** | 2.5 | 2.5 | 4 | **1** |
| **manufacturing email** | **1** | 3 | 2 | 4 |
| **littlerock foodweb** | **1** | 4 | 2 | 3 |
| **jazz** | **1** | 2 | 3 | 4 |
| **residence hall friends** | **1** | 2.5 | 4 | 2.5 |
| **haggle contacts** | **1.5** | 3.5 | **1.5** | 3.5 |
| **worm nervoussys** | **1** | 3 | 2 | 4 |
| **netsci** | 2 | **1** | 4 | 3 |
| **infectious contacts** | **1** | 2 | 4 | 3 |
| **flightmap** | **1** | 4 | 2 | 3 |
| **email** | **2** | **2** | 4 | **2** |
| **polblog** | **1** | 3.5 | 2 | 3.5 |
| **mean ranking** | **1.89** | 2.50 | 2.71 | 2.89 |
| **mean ranking (without littlerock foodweb)** | **1.94** | 2.42 | 2.75 | 2.89 |

**Table 4. Permutation test for the mean ranking in link-prediction on small-size real networks.**
For each pair of methods, a permutation test for the mean has been applied to the two vectors of link-prediction rankings on the real networks (columns of Table 3), using 10000 iterations. The table reports the pairwise p-values, adjusted for multiple hypothesis comparison by the Benjamini–Hochberg correction. The p-values lower than the significance level of 0.05 are highlighted in bold.

| p-value | SPM | CH | SBM | FBM |
|---|---|---|---|---|
| **SPM** |  | 0.064 | **0.029** | **0.014** |
| **CH** | 0.064 |  | 0.275 | 0.160 |
| **SBM** | **0.029** | 0.275 |  | 0.275 |
| **FBM** | **0.014** | 0.160 | 0.275 |  |

**Table 5. Computational time on small-size real networks.**
For each network 10% of links have been randomly removed (10 iterations for SBM due to the high computational time, 100 iterations for the other methods) and the algorithms have been executed in order to assign likelihood scores to the non-observed links in these reduced networks. The table reports for each network the mean computational time (in seconds) over the random iterations and the mean time over the entire dataset. The networks are sorted by increasing number of nodes $N$. Note that SPM, CH and FBM are MATLAB implementations, whereas SBM is a C library. The methods have been run in the same workstation, see the section Hardware and Software for further details. For each network the best method (or methods) is highlighted in bold.

|  | SPM | CH | FBM | SBM |
|---|---|---|---|---|
| mouse neural | **0.1** | 0.2 | **0.1** | 6.5 |
| karate | **0.1** | 0.2 | **0.1** | 5.0 |
| dolphins | **0.1** | 0.3 | **0.1** | 28.0 |
| macaque neural | **0.1** | 0.3 | **0.1** | 182.0 |
| polbooks | **0.1** | 0.4 | 0.2 | 135.9 |
| ACM2009 contacts | **0.1** | 0.4 | 0.2 | 387.2 |
| football | **0.1** | 0.4 | 0.2 | 123.0 |
| physicians innovation | **0.1** | 0.5 | 0.2 | 152.3 |
| manufacturing email | **0.1** | 0.7 | 0.2 | 1401.3 |
| littlerock foodweb | **0.1** | 0.8 | 0.2 | 1526.7 |
| jazz | **0.1** | 0.9 | 0.3 | 874.0 |
| residence hall friends | **0.1** | 1.1 | 0.4 | 1559.8 |
| haggle contacts | **0.2** | 1.6 | **0.2** | 1533.9 |
| worm nervoussys | **0.2** | 1.9 | 0.8 | 2199.0 |
| netsci | **0.3** | 3.0 | 1.4 | 2749.0 |
| infectious contacts | **0.3** | 3.3 | 2.1 | 8526.5 |
| flightmap | 8.0 | 4.6 | **2.5** | 91873.0 |
| email | **4.6** | 24.3 | 47.5 | 103380.0 |
| polblog | **7.2** | 27.9 | 89.0 | 172284.0 |
| mean time (s) | **1.2** | 3.8 | 7.7 | 20469.8 |

**Table 6. Permutation test for the mean ranking in link-prediction on PSO networks.**
For each parameter combination (*m, T, N*) of the PSO networks shown in Fig. 3, the methods are ranked by decreasing mean precision (over the random iterations), considering an average rank in case of ties. The mean ranking of the methods over all the parameter combinations represents the final evaluation for a proper comparison of the performance and it is shown in the right side of the table. For each pair of methods, a permutation test for the mean has been applied to the two vectors of link-prediction rankings on the PSO networks, using 10000 iterations. The table reports the pairwise p-values, adjusted for multiple hypothesis comparison by the Benjamini–Hochberg correction. The p-values lower than the significance level of 0.05 are highlighted in bold.

| p-value | CH | SPM | FBM | SBM | mean ranking |
|---|---|---|---|---|---|
| **CH** | | **< 0.001** | **< 0.001** | **< 0.001** | **1.30** |
| **SPM** | **< 0.001** | | **< 0.001** | **< 0.001** | 1.78 |
| **FBM** | **< 0.001** | **< 0.001** | | 0.131 | 3.39 |
| **SBM** | **< 0.001** | **< 0.001** | 0.131 | | 3.54 |

**Table 7. Permutation test for the mean ranking in link-prediction on nonuniform PSO networks with 8 communities.**

For each parameter combination (*m, T, N*) of the nPSO networks shown in Fig. 4, the methods are ranked by decreasing mean precision (over the random iterations), considering an average rank in case of ties. The mean ranking of the methods over all the parameter combinations represents the final evaluation for a proper comparison of the performance and it is shown in the right side of the table. For each pair of methods, a permutation test for the mean has been applied to the two vectors of link-prediction rankings on the PSO networks, using 10000 iterations. The table reports the pairwise p-values, adjusted for multiple hypothesis comparison by the Benjamini–Hochberg correction. The p-values lower than the significance level of 0.05 are highlighted in bold.

| p-value | CH | SPM | SBM | FBM | mean ranking |
|---|---|---|---|---|---|
| **CH** | | **0.024** | **< 0.001** | **< 0.001** | **1.69** |
| **SPM** | **0.024** | | **0.001** | **< 0.001** | 1.96 |
| **SBM** | **< 0.001** | **0.001** | | **< 0.001** | 2.59 |
| **FBM** | **< 0.001** | **< 0.001** | **< 0.001** | | 3.76 |

**Table 8. Statistics of AS Internet snapshots and large-size real networks.**
The first half of the table reports the statistics for the AS Internet snapshots, whereas the second half the large-size real networks. Note that also the last AS Internet snapshot has been considered in the simulations on the large-size real networks. For each network several statistics have been computed. $N$ is the number of nodes. $E$ is the number of edges. The parameter $m$, as in the PSO model, refers to half of the average node degree. $D$ is the network density. $C$ is the average clustering coefficient, computed for each node as the number of links between its neighbours over the number of possible links [55]. $L$ is the characteristic path length of the network [55]. *LCP-corr* is the Local-Community-Paradigm correlation [3], representing the correlation between the number of common-neighbours and the number of links between them, looking at each pair of connected nodes in the network. *Struct-cons* is the structural consistency [9], a quantitative index that estimates the link predictability of the network. Power-law is the exponent γ of the power-law distribution estimated from the observed degree distribution of the network using the maximum likelihood procedure described in [33].

|  | N | E | m | D | C | L | LCP corr | Struct cons | Power law |
|---|---|---|---|---|---|---|---|---|---|
| **ARK200909** | 24091 | 59531 | 2.47 | 0.0039 | 0.36 | 3.53 | 0.95 | 0.10 | 2.12 |
| **ARK200912** | 25910 | 63435 | 2.45 | 0.0031 | 0.36 | 3.54 | 0.94 | 0.10 | 2.11 |
| **ARK201003** | 26307 | 66089 | 2.51 | 0.0012 | 0.37 | 3.53 | 0.94 | 0.10 | 2.26 |
| **ARK201006** | 26756 | 68150 | 2.55 | 0.0011 | 0.37 | 3.51 | 0.95 | 0.09 | 2.08 |
| **ARK201009** | 28353 | 73722 | 2.60 | 0.0009 | 0.37 | 3.52 | 0.94 | 0.10 | 2.23 |
| **ARK201012** | 29333 | 78054 | 2.66 | 0.0002 | 0.38 | 3.50 | 0.95 | 0.10 | 2.22 |
| **odlis** | 2898 | 16376 | 5.65 | 0.0002 | 0.30 | 3.17 | 0.93 | 0.10 | 2.63 |
| **advogato** | 5042 | 39227 | 7.78 | 0.0002 | 0.25 | 3.27 | 0.90 | 0.16 | 2.73 |
| **arxiv astroph** | 17903 | 196972 | 11.00 | 0.0002 | 0.63 | 4.19 | 0.95 | 0.67 | 2.83 |
| **thesaurus** | 23132 | 297094 | 12.84 | 0.0002 | 0.09 | 3.49 | 0.87 | 0.07 | 2.84 |
| **arxiv hepth** | 27400 | 352021 | 12.85 | 0.0002 | 0.31 | 4.28 | 0.92 | 0.27 | 2.86 |
| **facebook** | 43953 | 182384 | 4.15 | 0.0002 | 0.11 | 5.60 | 0.87 | 0.09 | 3.66 |

**Table 9. Precision evaluation of link-prediction in time on AS Internet networks.**
Six AS Internet network snapshots are available from September 2009 to December 2010, at time steps of 3 months. For every snapshot at times $i = [1, 5]$ the algorithms have been executed in order to assign likelihood scores to the non-observed links and the link-prediction performance has been evaluated computing the precision with respect to every future time point $j = [i+1, 6]$. Considering a pair of time points $(i, j)$, the non-observed links at time $i$ are ranked by decreasing likelihood scores and the precision is computed as the percentage of links that appear at time $j$ among the top-$r$ in the ranking, where $r$ is the total number of non-observed links at time $i$ that appear at time $j$. Non-observed links at time $i$ involving nodes that disappear at time $j$ are not considered in the ranking. The table reports for each method a 5-dimensional upper triangular matrix, containing as element $(i, j)$ the precision of the link-prediction from time $i$ to time $j+1$. On the right side, the methods are ranked by the mean precision computed over all the time combinations. The last column shows the time required for executing the methods on the biggest network (last snapshot, December 2010), after the removal of 10% of the links, as average over 10 iterations. For each comparison the best method is highlighted in bold.

| CH | | | | | SPM | | | | | | mean precision | mean ranking | mean time |
|---|---|---|---|---|---|---|---|---|---|---|---|---|---|
| **0.11** | **0.12** | **0.13** | **0.14** | **0.14** | 0.08 | 0.09 | 0.09 | 0.10 | 0.11 | **CH** | **0.13** | **1** | **1.2 h** |
| | 0.12 | 0.13 | 0.14 | 0.14 | | 0.07 | 0.08 | 0.09 | 0.10 | SPM | 0.09 | 2 | 6.8 h |
| | | 0.12 | 0.13 | 0.14 | | | 0.08 | 0.09 | 0.10 | | | | |
| | | | 0.12 | 0.13 | | | | 0.08 | 0.09 | | | | |
| | | | | 0.12 | | | | | 0.09 | | | | |

**Table 10. Evaluation of link-prediction on large-size real networks.**
For each network 10% of links have been randomly removed (10 iterations) and the algorithms have been executed in order to assign likelihood scores to the non-observed links in these reduced networks. In order to evaluate the performance, the links are ranked by likelihood scores and the precision is computed as the percentage of removed links among the top-$r$ in the ranking, where $r$ is the total number of links removed. The table reports for each network the mean precision over the random iterations. The last rows show the mean precision over the entire dataset, the mean ranking computed as described in Table 3 and the p-value of the permutation test for the mean ranking applied as described in Table 4, without any correction since there is a single comparison. For each network the best method is highlighted in bold. The networks are sorted by increasing number of nodes $N$.

|  | CH | SPM |
|---|---|---|
| odlis | **0.12** | 0.08 |
| advogato | **0.16** | 0.15 |
| arxiv astroph | 0.53 | **0.67** |
| thesaurus | **0.08** | 0.07 |
| arxiv hepth | 0.22 | **0.27** |
| ARK201012 | **0.16** | 0.11 |
| facebook | **0.11** | 0.10 |
| mean precision | 0.20 | **0.21** |
| mean ranking | **1.29** | 1.71 |
| p-value | 0.013 | |

**Table 11. Computational time on large-size real networks.**
For each network 10% of links have been randomly removed (10 iterations) and the algorithms have been executed in order to assign likelihood scores to the non-observed links in these reduced networks. The table reports for each network the mean computational time over the random iterations and the mean time over the entire dataset. The networks are sorted by increasing number of nodes $N$. The methods have been run in the same workstation, see the section Hardware and software for further details. For each network the best method is highlighted in bold.

|  | CH | SPM |
|---|---|---|
| **odlis** | 0.7 min | **0.3 min** |
| **advogato** | 2.2 min | **1.6 min** |
| **arxiv astroph** | **21.9 min** | 70.8 min |
| **thesaurus** | **1.3 h** | 2.5 h |
| **arxiv hepth** | **1.1 h** | 3.9 h |
| **ARK201012** | **1.2 h** | 6.8 h |
| **facebook** | **2.1 h** | 15.3 h |
| **mean time** | **0.9 h** | 4.2 h |

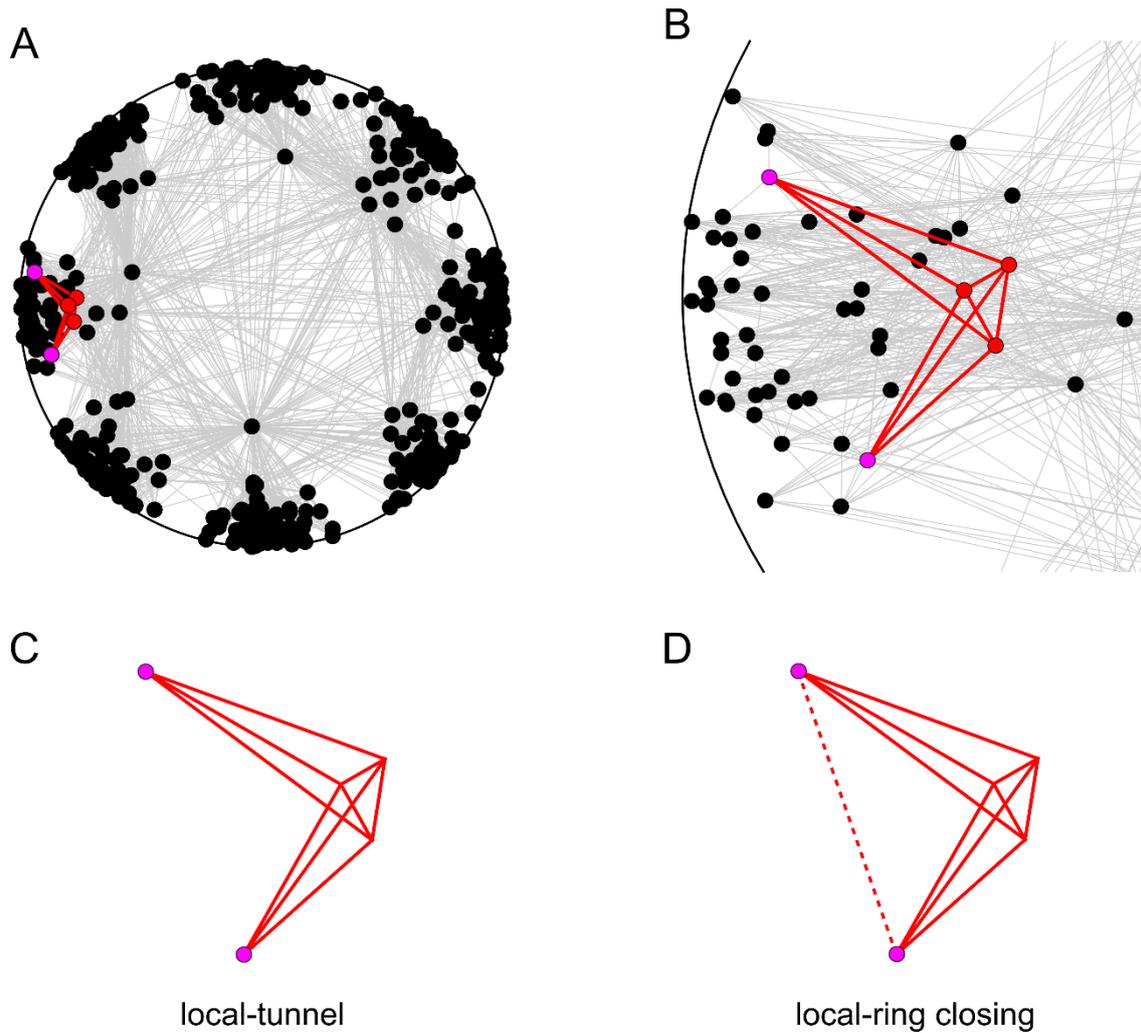

**Figure 1. Explanatory example of the local-tunnel and local-ring closing procedure.**
(A) Representation of a nPSO network ($\gamma = 3$, $m = 10$, $T = 0.1$, $N = 500$, $C = 8$) in the hyperbolic space. In violet two non-adjacent nodes, whereas in red their CNs, the links from the two nodes to their CNs and the LCLs. (B) A zoom of the network in the region of two non-adjacent nodes considered for link formation. (C) The local-tunnel is the ensemble of the local-paths and the LCLs, and provides a route of connectivity between the two non-adjacent nodes. (D) The addition of the link between the two non-adjacent nodes leads to the local-ring closing procedure, which transforms the local-tunnel in a local-ring.

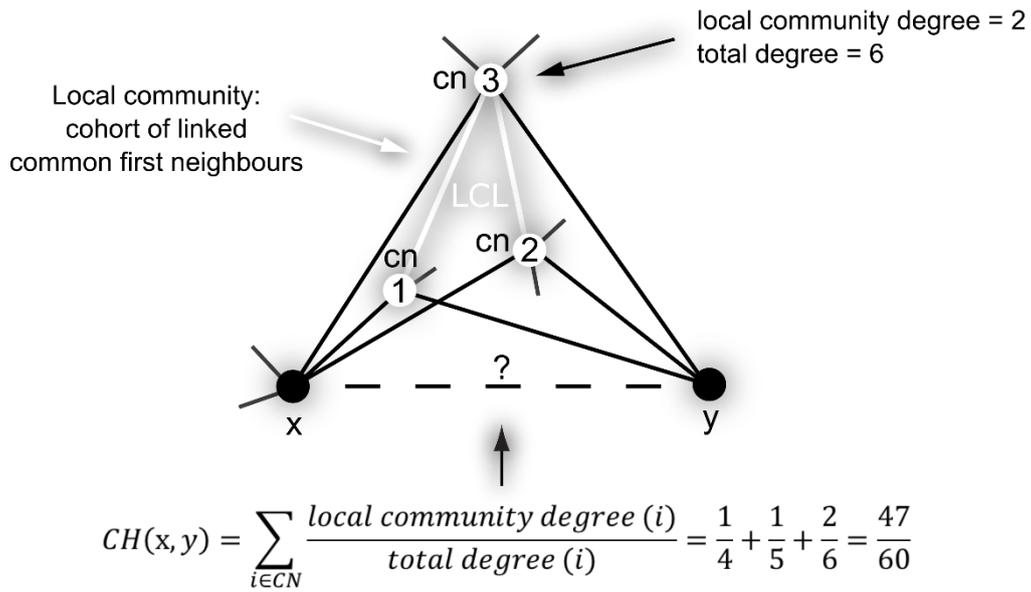

$$CH(x,y) = \sum_{i \in CN} \frac{local\ community\ degree\ (i)}{total\ degree\ (i)} = \frac{1}{4} + \frac{1}{5} + \frac{2}{6} = \frac{47}{60}$$

**Figure 2. Example of link-prediction using the Cannistraci-Hebb network automaton model.** The figure shows an explanatory example for the topological link-prediction performed using the Cannistraci-Hebb (CH) rule. The two black nodes represent the seed nodes whose non-observed interaction should be scored with a likelihood. The three white nodes are the common-neighbours (CNs) of the seed nodes, further neighbours are not shown for simplicity. The cohort of linked common-neighbours forms the local-community and the links between them are called local-community-links (LCLs). Neighbours of the CNs that are neither the seed nodes nor in the local-community are indicated through chunks of outgoing links, which we named eLCLs. The mathematical formula for the CH index is reported, together with the detailed steps for computing the likelihood score for the link under analysis.

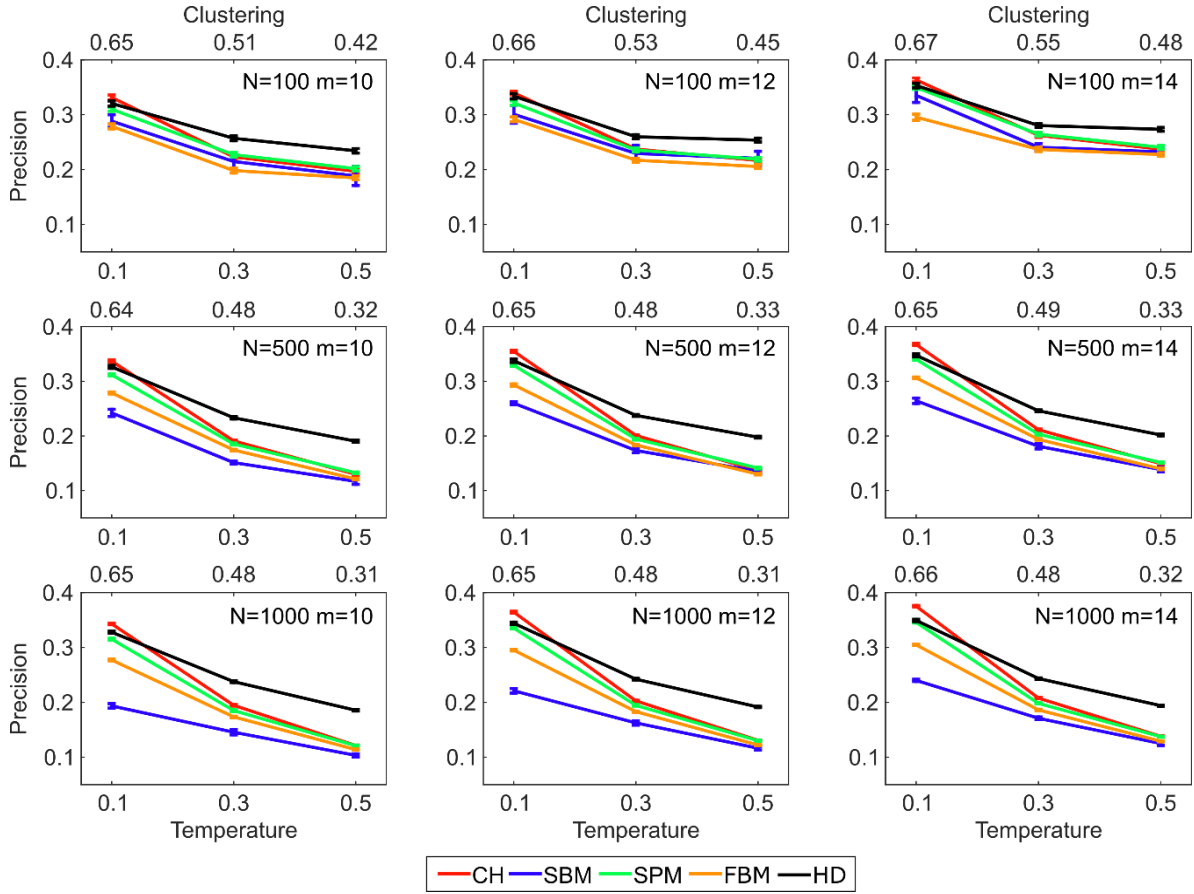

**Figure 3. Precision evaluation of link-prediction on PSO networks.**
Synthetic networks have been generated using the PSO model with parameters $\gamma = 3$ (power-law degree distribution exponent), $m = [10, 12, 14]$ (half of average degree), $T = [0.1, 0.3, 0.5]$ (temperature, inversely related to the clustering coefficient) and $N = [100, 500, 1000]$ (network size). The values chosen for the parameter $m$ are centered around the average $m$ computed on the dataset of small-size real networks. The values chosen for $N$ and $T$ are intended to cover the range of network size and clustering coefficient observed in the dataset of small-size real networks. Since the average $\gamma$ estimated on the dataset of small-size real networks is higher than the typical range $2 < \gamma < 3$ [33], we choose $\gamma = 3$. For each combination of parameters, 100 networks have been generated. For each network 10% of links have been randomly removed and the algorithms have been executed in order to assign likelihood scores to the non-observed links in these reduced networks. In order to evaluate the performance, the links are ranked by likelihood scores and the precision is computed as the percentage of removed links among the top-$r$ in the ranking, where $r$ is the total number of links removed. The plots report for each parameter combination the mean precision and standard error over the random iterations. Note that for SBM only 10 networks have been considered due to the high computational time.

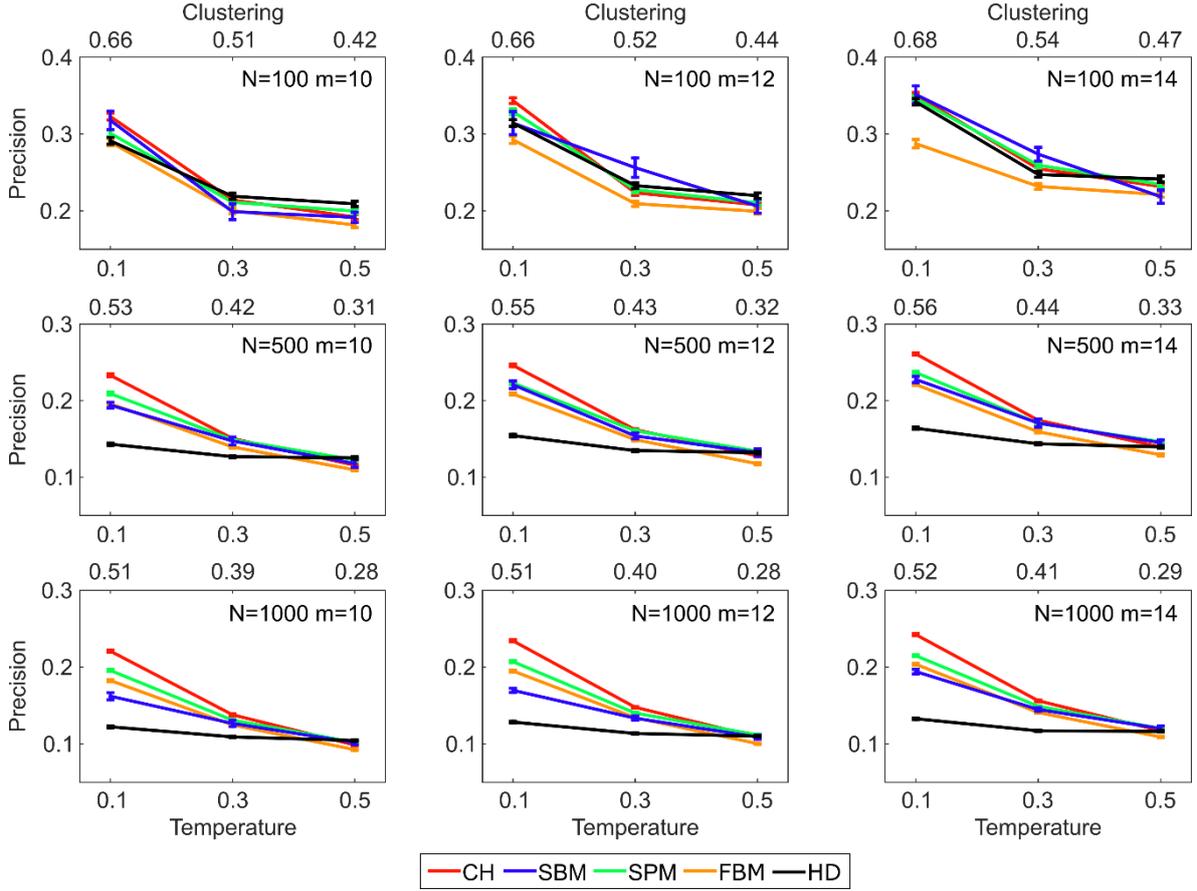

**Figure 4. Precision evaluation of link-prediction on nPSO networks with 8 communities.**
Synthetic networks have been generated using the nonuniform PSO model with parameters $\gamma = 3$ (power-law degree distribution exponent), $m = [10, 12, 14]$ (half of average degree), $T = [0.1, 0.3, 0.5]$ (temperature, inversely related to the clustering coefficient), $N = [100, 500, 1000]$ (network size) and 8 communities. The values chosen for the parameter $m$ are centered around the average $m$ computed on the dataset of small-size real networks. The values chosen for $N$ and $T$ are intended to cover the range of network size and clustering coefficient observed in the dataset of small-size real networks. Since the average $\gamma$ estimated on the dataset of small-size real networks is higher than the typical range $2 < \gamma < 3$ [33], we choose $\gamma = 3$. For each combination of parameters, 100 networks have been generated. For each network 10% of links have been randomly removed and the algorithms have been executed in order to assign likelihood scores to the non-observed links in these reduced networks. In order to evaluate the performance, the links are ranked by likelihood scores and the precision is computed as the percentage of removed links among the top-$r$ in the ranking, where $r$ is the total number of links removed. The plots report for each parameter combination the mean precision and standard error over the random iterations. Note that for SBM only 10 networks have been considered due to the high computational time.

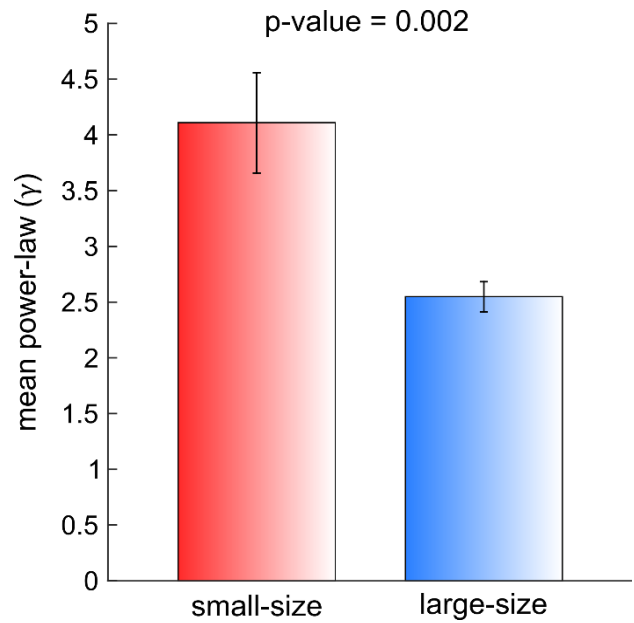

**Figure 5. Comparison of power-law exponent between small-size and large-size real networks.**
The barplot reports the mean and standard error of the power-law exponent $\gamma$ estimated from the observed degree distribution of the small-size and large-size real networks. A permutation test for the mean (10000 iterations) has been applied to the two vectors of power-law exponents (rightmost columns of Table 1 and Table 8) and the p-value is shown on top of the barplot.

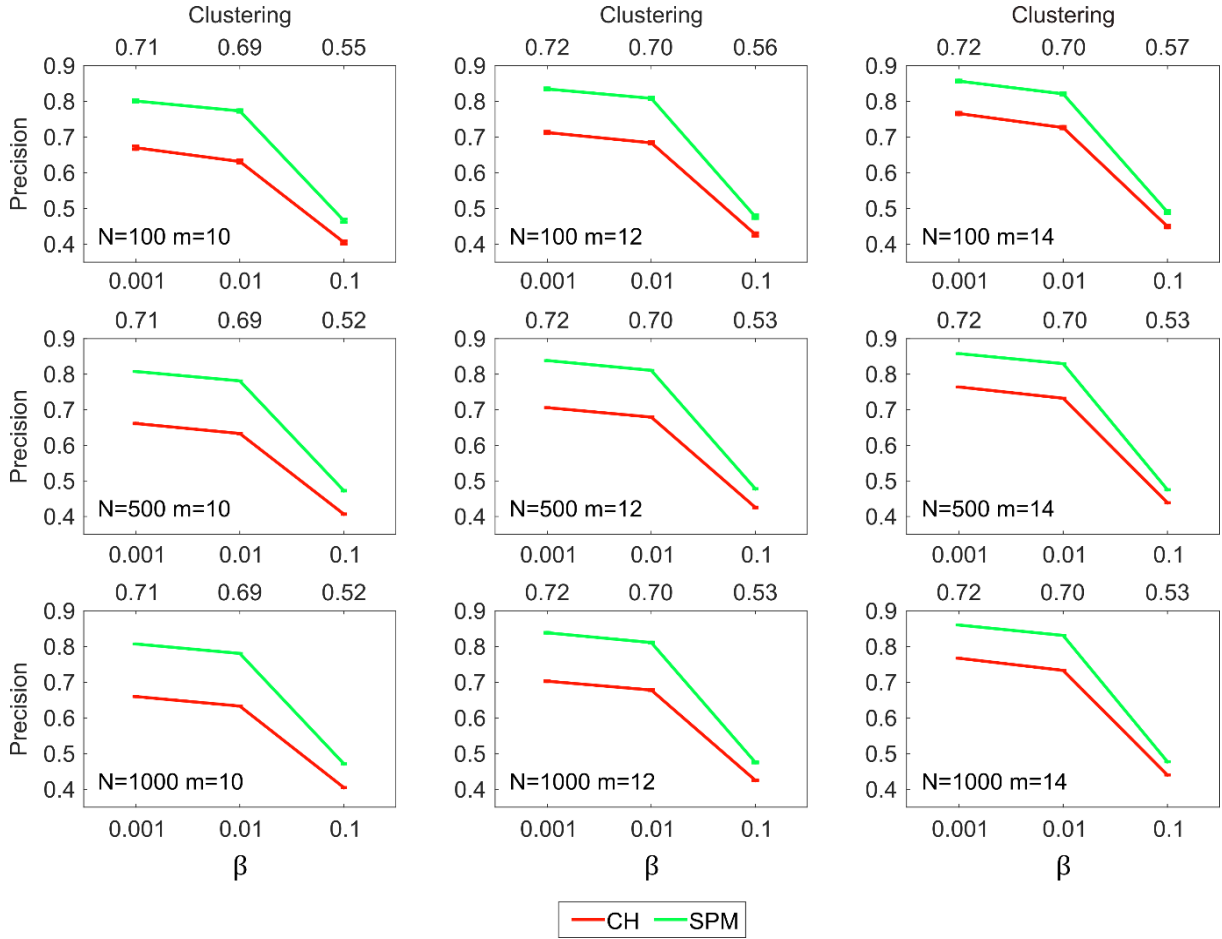

**Figure 6. Precision evaluation of link-prediction on Watts-Strogatz networks.**
Synthetic networks have been generated using the Watts-Strogatz model with parameters $N$ = [100, 500, 1000] (network size), $m$ = [10, 12, 14] (half of average degree) and $\beta$ = [0.001, 0.01, 0.1] (rewiring probability). The values chosen for the parameters $N$ and $m$ are the same used for the synthetic networks generated using the PSO model. The values chosen for $\beta$ are intended to produce networks with different properties mainly in terms of clustering coefficient and characteristic path length, as expected according to the Watts-Strogatz model. Low values of $\beta$ generate strong clustering. For each combination of parameters, 100 networks have been generated. For each network 10% of links have been randomly removed and the algorithms have been executed in order to assign likelihood scores to the non-observed links in these reduced networks. In order to evaluate the performance, the links are ranked by likelihood scores and the precision is computed as the percentage of removed links among the top-$r$ in the ranking, where $r$ is the total number of links removed. The plots report for each parameter combination the mean precision and standard error over the random iterations. A statistical test between the performances of the two methods is not needed since the mean of SPM is always higher and the standard error is negligible, practically the p-value is theoretically zero.

# SUPPLEMENTARY INFORMATION

**Suppl. Table 1. Link prediction in time on AS Internet networks: CH versus RA.**
Six AS Internet network snapshots are available from September 2009 to December 2010, at time steps of 3 months. For every snapshot at times $i = [1, 5]$ the algorithms have been executed in order to assign likelihood scores to the non-observed links and the link prediction performance has been evaluated computing the precision with respect to every future time point $j = [i+1, 6]$. Considering a pair of time points $(i, j)$, the non-observed links at time $i$ are ranked by decreasing likelihood scores and the precision is computed as the percentage of links that appear at time $j$ among the top-$r$ in the ranking, where $r$ is the total number of non-observed links at time $i$ that appear at time $j$. Non-observed links at time $i$ involving nodes that disappear at time $j$ are not considered in the ranking. The table reports for each method a 5-dimensional upper triangular matrix, containing as element $(i, j)$ the precision of the link prediction from time $i$ to time $j+1$. On the right side, the methods are ranked by the mean precision computed over all the time combinations, the mean ranking is also shown. For each comparison the best method is highlighted in bold.

| CH | | | | | RA | | | | | | mean precision | mean ranking |
|---|---|---|---|---|---|---|---|---|---|---|---|---|
| **0.11** | **0.12** | **0.13** | **0.14** | **0.14** | 0.10 | 0.11 | 0.11 | 0.12 | 0.12 | CH | 0.13 | 1 |
| | **0.12** | **0.13** | **0.14** | **0.14** | | 0.09 | 0.10 | 0.11 | 0.12 | RA | 0.11 | 2 |
| | | **0.12** | **0.13** | **0.14** | | | 0.09 | 0.11 | 0.12 | | | |
| | | | **0.12** | **0.13** | | | | 0.10 | 0.11 | | | |
| | | | | **0.12** | | | | | 0.10 | | | |

**Suppl. Table 2. Evaluation of link prediction on large-size real networks: CH versus RA.**
For each network 10% of links have been randomly removed (10 iterations) and the algorithms have been executed in order to assign likelihood scores to the non-observed links in these reduced networks. In order to evaluate the performance, the links are ranked by likelihood scores and the precision is computed as the percentage of removed links among the top-$r$ in the ranking, where $r$ is the total number of links removed. The table reports for each network the mean precision over the random iterations. The last rows show the mean precision over the entire dataset, the mean ranking computed as described in Table 3 and the p-value of the permutation test for the mean ranking applied as described in Table 4, without any correction since there is a single comparison. For each network the best method is highlighted in bold. The networks are sorted by increasing number of nodes $N$.

|  | CH | RA |
|---|---|---|
| **odlis** | **0.12** | 0.10 |
| **advogato** | **0.16** | 0.14 |
| **arxiv astroph** | 0.53 | **0.64** |
| **thesaurus** | **0.08** | 0.03 |
| **arxiv hepth** | **0.22** | 0.20 |
| **ARK201012** | **0.16** | **0.16** |
| **facebook** | **0.11** | 0.06 |
| mean precision | **0.20** | 0.19 |
| mean ranking | **1.21** | 1.79 |
| p-value | **0.004** | |

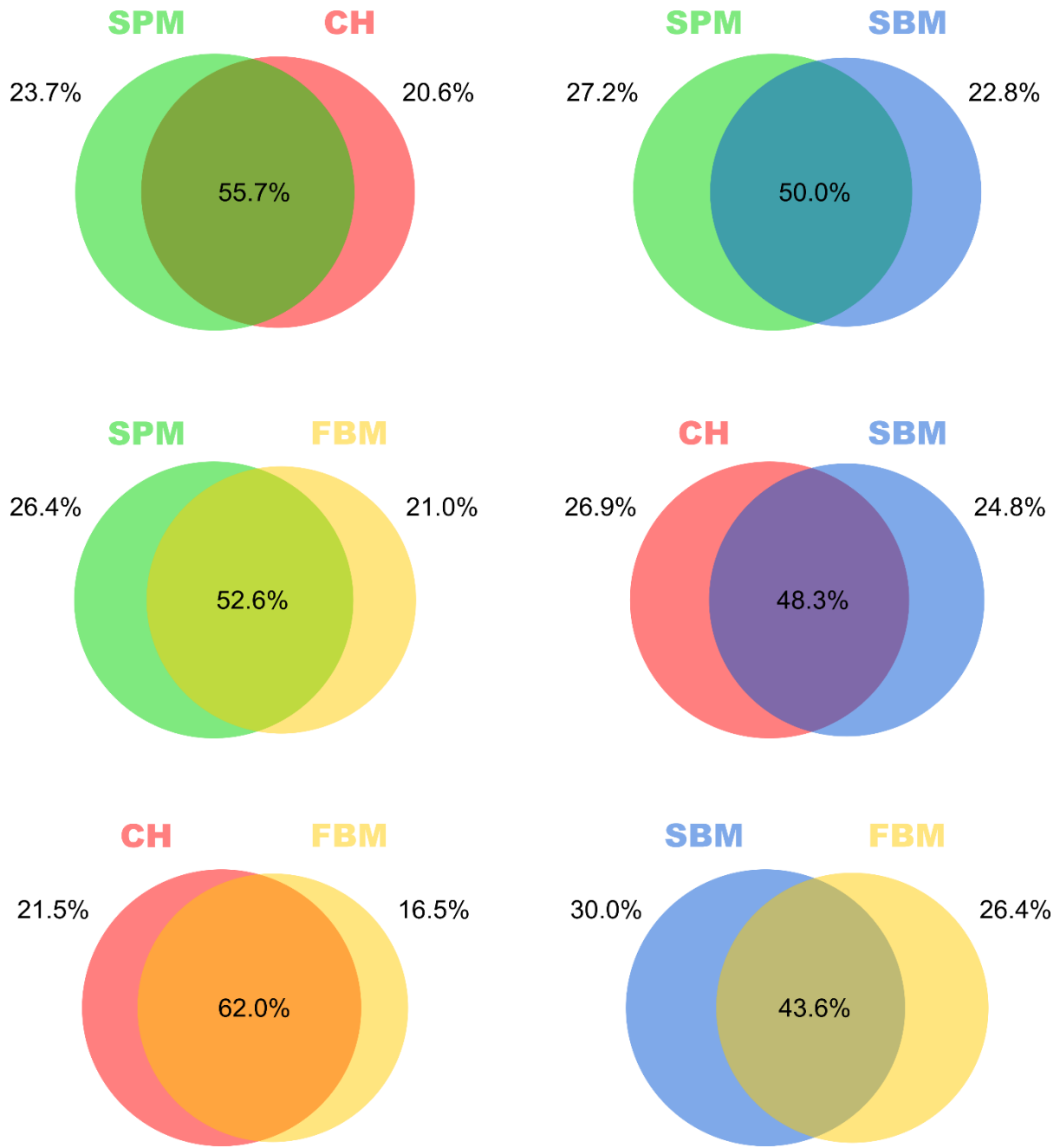

**Suppl. Figure 1. Pairwise Venn diagrams of correctly predicted links on small-size real networks.** For each pair of methods, the overlap of the correctly predicted links has been analysed. For each small-size real network (and for each of 10 iterations), considering the entire set of links that have been correctly predicted by two methods, the percentage of these links that are shared or not is computed and reported in the corresponding Venn diagram, as average over all the networks and iterations.

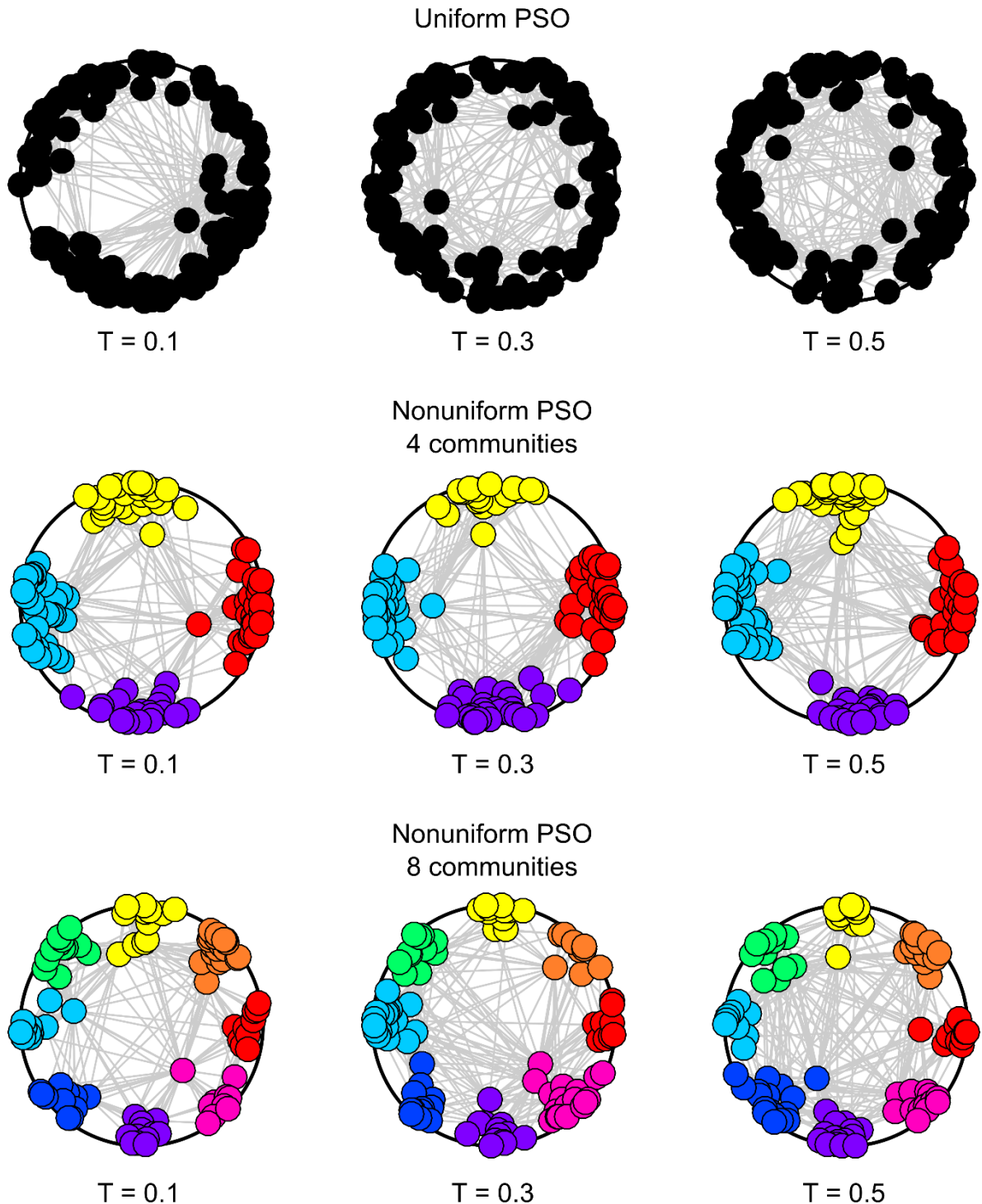

**Suppl. Figure 2. Uniform versus nonuniform PSO model.**
Synthetic networks have been generated using both the uniform and nonuniform PSO model (4 and 8 communities) with parameters $\gamma = 3$ (power-law degree distribution exponent), $m = 5$ (half of average degree), $T = [0.1, 0.3, 0.5]$ (temperature, inversely related to the clustering coefficient) and $N = 100$ (network size). The plots show for each parameter combination a representation in the hyperbolic space of the networks. When generated by the nonuniform PSO model, the nodes are coloured according to the communities.

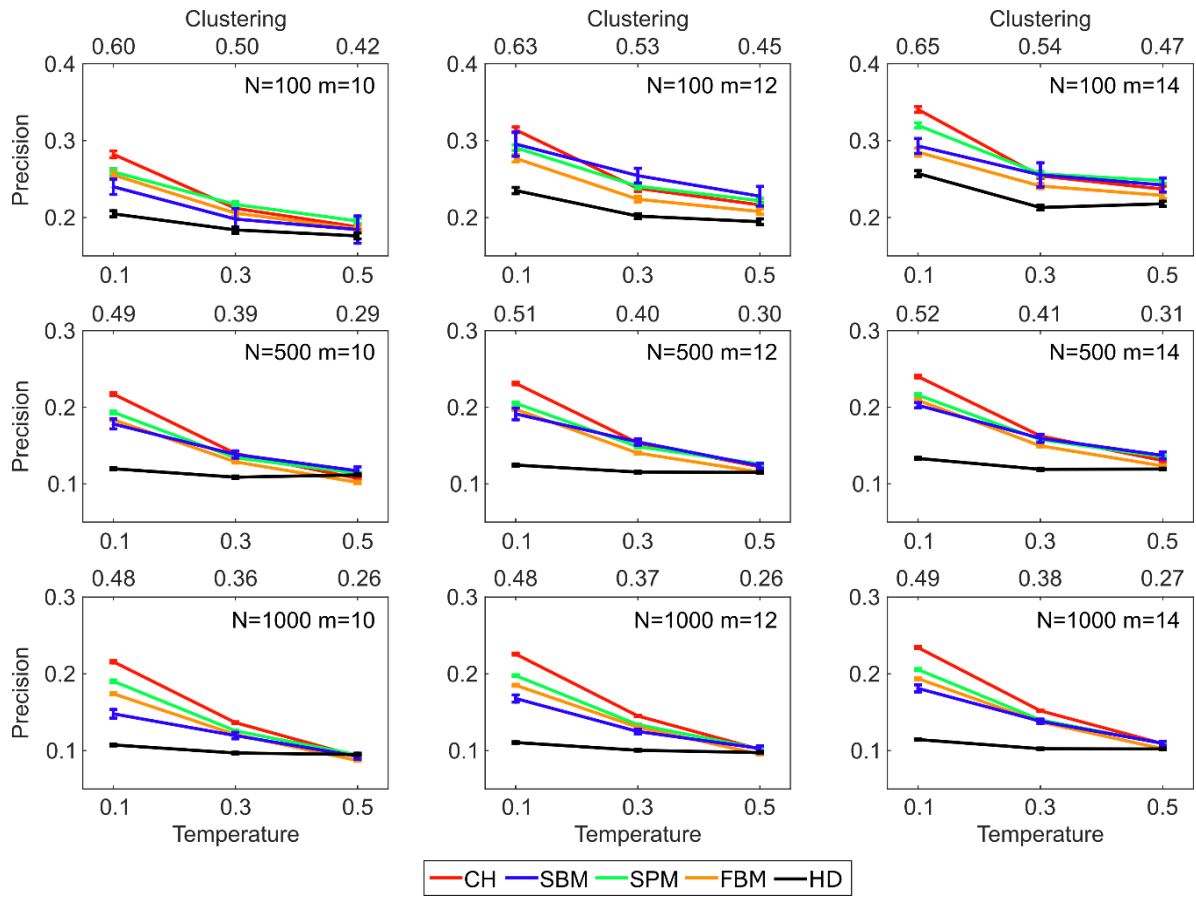

**Suppl. Figure 3. Precision evaluation of link prediction on nPSO networks with 4 communities.**
Synthetic networks have been generated using the nonuniform PSO model with parameters $\gamma = 3$ (power-law degree distribution exponent), $m = [10, 12, 14]$ (half of average degree), $T = [0.1, 0.3, 0.5]$ (temperature, inversely related to the clustering coefficient), $N = [100, 500, 1000]$ (network size) and 4 communities. The values chosen for the parameter $m$ are centered around the average $m$ computed on the dataset of small-size real networks. The values chosen for $N$ and $T$ are intended to cover the range of network size and clustering coefficient observed in the dataset of small-size real networks. Since the average $\gamma$ estimated on the dataset of small-size real networks is higher than the typical range $2 < \gamma < 3$, we choose $\gamma = 3$. For each combination of parameters, 100 networks have been generated. For each network 10% of links have been randomly removed and the algorithms have been executed in order to assign likelihood scores to the non-observed links in these reduced networks. In order to evaluate the performance, the links are ranked by likelihood scores and the precision is computed as the percentage of removed links among the top-$r$ in the ranking, where $r$ is the total number of links removed. The plots report for each parameter combination the mean precision and standard error over the random iterations. Note that for SBM only 10 networks have been considered due to the high computational time.

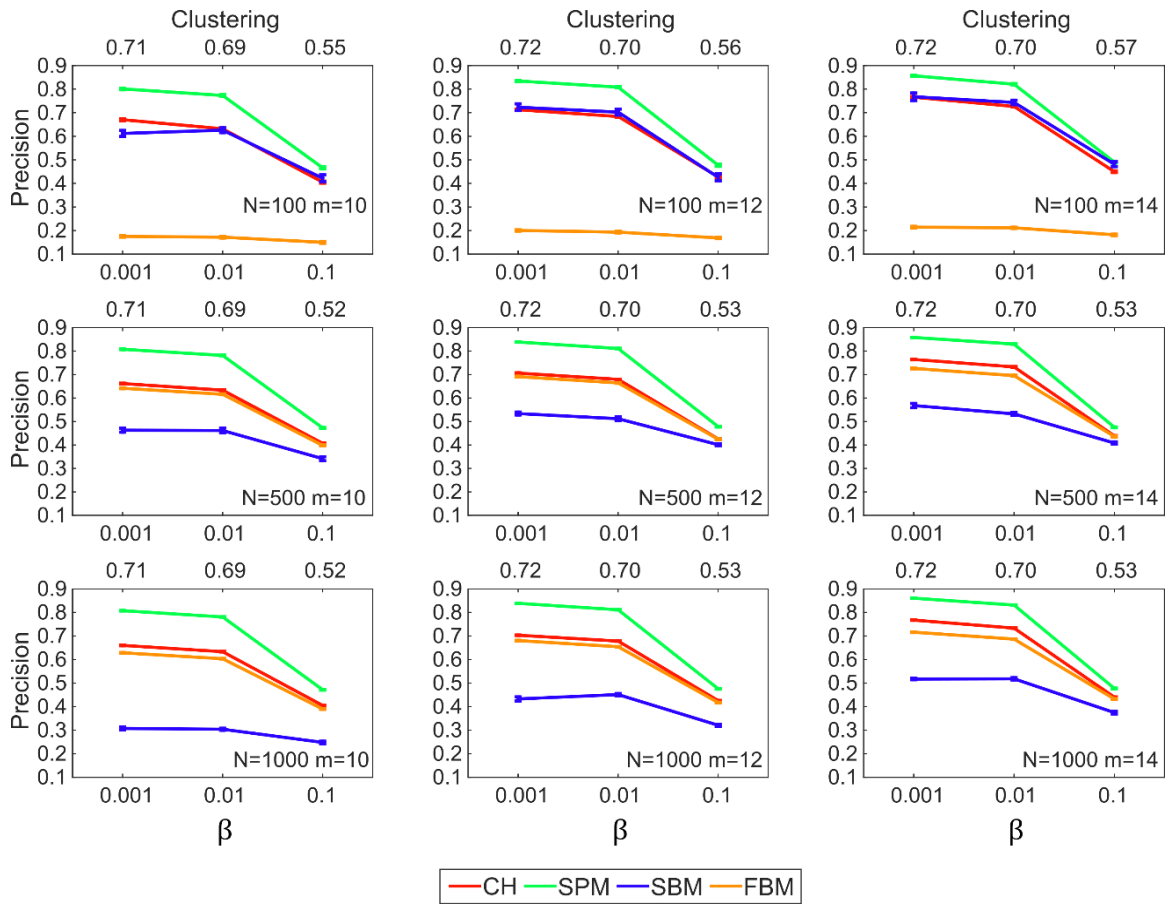

**Suppl. Figure 4. Precision evaluation of link prediction on Watts-Strogatz networks.**
Synthetic networks have been generated using the Watts-Strogatz model with parameters $N$ = [100, 500, 1000] (network size), $m$ = [10, 12, 14] (half of average degree) and $\beta$ = [0.001, 0.01, 0.1] (rewiring probability). The values chosen for the parameters $N$ and $m$ are the same used for the synthetic networks generated using the PSO model. The values chosen for $\beta$ are intended to produce networks with different properties mainly in terms of clustering coefficient and characteristic path length, as expected according to the Watts-Strogatz model. Low values of $\beta$ generate strong clustering. For each combination of parameters, 100 networks have been generated. For each network 10% of links have been randomly removed and the algorithms have been executed in order to assign likelihood scores to the non-observed links in these reduced networks. In order to evaluate the performance, the links are ranked by likelihood scores and the precision is computed as the percentage of removed links among the top-$r$ in the ranking, where $r$ is the total number of links removed. The plots report for each parameter combination the mean precision and standard error over the random iterations. Note that for SBM only 10 networks have been considered due to the high computational time.